\documentclass[aps,prd,10pt,twocolumn,superscriptaddress,floatfix,nofootinbib,amsmath,amssymb,altaffilletter,preprintnumbers,eprint,nolongbibliography]{revtex4-2}

\usepackage[normalem]{ulem}
\usepackage{natbib}
\usepackage[english]{babel}
\usepackage[utf8]{inputenc}

\usepackage{enumitem}
\usepackage{braket}
\usepackage{footnote}
\usepackage{tablefootnote}
\usepackage{amsmath,amssymb,amsfonts,mathrsfs}
\usepackage{booktabs}
\usepackage{mathtools}
\usepackage{bm}
\usepackage{dcolumn}
\usepackage{graphicx}   
\usepackage{latexsym}
\usepackage[acronym]{glossaries}
\usepackage{physics}
\usepackage{siunitx}
\usepackage{comment}

\usepackage{tikz}
\usetikzlibrary{calc}
\usetikzlibrary{decorations.pathreplacing, snakes}
\usepackage{adjustbox}

\DeclareSIUnit{\persqrthz}{\ensuremath{\text{Hz}^{-1/2}}}
\DeclareMathOperator*{\dprime}{\prime \prime}

\usepackage[svgnames]{xcolor}
\usepackage[final]{hyperref}
\hypersetup{%
    colorlinks=true,
    citecolor=Blue,
    linkcolor=Blue,
    urlcolor=Blue
    }%
\newcommand{\fo}{{f_\mathrm{orb}}}
\newcommand{\oo}{{\omega_\mathrm{orb}}}
\newcommand{\og}{{\omega_\mathrm{GW}}}
\newcommand{\fg}{{f_\mathrm{GW}}}
\newcommand{\osc}{{\sim \mathbf{os}_{f_\mathrm{GW}}}}

\newcommand{\myhyperref}[1]{\hyperref[#1]{\ref{#1}}}
\usepackage{academicons}
\definecolor{orcidlogocol}{HTML}{A6CE39}
\newcommand{\orcid}[1]{\href{https://orcid.org/#1}{\textcolor[HTML]{A6CE39}{\aiOrcid}}}

\begin{document}

\title{The Doppler boosted LISA response to gravitational waves \vspace{-.2cm}}

\author{Tom van der Steen \orcid{0009-0002-0910-9072}}
\email{tom.vandersteen@kuleuven.be}
\affiliation{Institute for Theoretical Physics, KU Leuven,
Celestijnenlaan 200D, B-3001 Leuven, Belgium}
\affiliation{Leuven Gravity Institute, KU Leuven,
Celestijnenlaan 200D box 2415, 3001 Leuven, Belgium}

\author{Henri Inchauspé \orcid{0000-0002-4664-6451}}
\email{henri.inchauspe@kuleuven.be}
\affiliation{Institute for Theoretical Physics, KU Leuven,
Celestijnenlaan 200D, B-3001 Leuven, Belgium}
\affiliation{Leuven Gravity Institute, KU Leuven,
Celestijnenlaan 200D box 2415, 3001 Leuven, Belgium}

\author{Thomas Hertog \orcid{0000-0002-9021-5966}}
\affiliation{Institute for Theoretical Physics, KU Leuven,
Celestijnenlaan 200D, B-3001 Leuven, Belgium}
\affiliation{Leuven Gravity Institute, KU Leuven,
Celestijnenlaan 200D box 2415, 3001 Leuven, Belgium}

\author{Aur\'elien Hees \orcid{0000-0002-2186-644X}}
\affiliation{LTE, Observatoire de Paris, Universit\'e PSL, Sorbonne Universit\'e, Universit\'e de Lille, LNE, CNRS 61 Avenue de l’Observatoire, 75014 Paris, France}

\begin{abstract}
Future space-based gravitational wave observatories like LISA, TianQin and Taji are expected to detect massive black hole binaries (MBHBs) with high signal-to-noise ratios (SNRs), ranging up to thousands. Such high-precision observations require accurate modeling of the detector response. However, current derivations of the response function neglect the motion of the spacecraft during light travel time, omitting velocity-dependent terms of order $\beta = v/c \sim 10^{-4}$. In this work, we derive the velocity-dependent corrections to the gravitational wave response. Focusing on LISA, we analyze the contribution of the velocity-terms for MBHBs in the mass range $[10^6,10^8]\:\mathrm{M}_{\odot}$ using a modified version of the state-of-the-art response simulator \texttt{lisagwresponse}. We find that corrections introduce residual SNRs up to $\sim 2$ for the loudest events and fractional differences up to $0.02\%$, compared to \texttt{lisagwresponse}. While small, these effects are comparable to current waveform modeling uncertainties and imprint distinctive sky-localization signatures, making them potentially relevant for parameter estimation of high-mass MBHBs and simulation of mock datasets.
\end{abstract}

\maketitle

\section{\label{Intro}Introduction}

The Laser Interferometer Space Antenna (LISA) is a space-based gravitational wave (GW) observatory planned for launch in the next decade \cite{amaro-seoaneLaserInterferometerSpace2017, colpiLISADefinitionStudy2024}. It will be sensitive to GWs in the millihertz regime, a frequency band inaccessible to ground-based detectors. This low-frequency window is rich in astrophysical sources, among which massive black hole binaries (MBHBs) are expected to be the loudest. With signal-to-noise ratios (SNRs) potentially reaching into the thousands, their signals will be unmistakable and overpower both noise and other signals \cite{colpiLISADefinitionStudy2024}. These sources offer a unique opportunity to test the nature of black holes (BHs) and to probe the validity of General Relativity (GR) in the strong-field regime with unprecedented precision \cite{bertiGravitationalwaveSpectroscopyMassive2006, bertiSpectroscopyKerrBlack2016, bhagwatLandscapeMassiveBlackhole2022}. Extracting the full scientific potential from these sources requires highly accurate modeling, not only of the gravitational waveforms themselves but also of the instrument’s response to those signals. While we focus specifically on LISA, similar considerations may also be relevant for other planned millihertz space-based GW observatories, such as Taiji and TianQin \cite{luoTaijiProgramConcise2021, liGravitationalWaveAstronomy2025}. Improving the accuracy of the LISA response by including spacecraft-velocity-dependent terms is the focus of this work.

LISA is a constellation of three spacecraft, each containing two free-falling test masses. GWs are detected by exchanging laser beams between the test masses and measuring the GW-induced phase shifts over the \SI{2.5e6}{\kilo\metre} arms. The calculation of the time delay and corresponding frequency shift induced by a GW on a photon traveling between two stationary test masses is well established in the literature \cite{estabrookResponseDopplerSpacecraft1975, burkeLargeScaleRandomGravitational1975, wahlquistDopplerResponseGravitational1987, cornishLISAResponseFunction2003, rakhmanovResponseTestMasses2005, finnResponseInterferometricGravitational2009, cornishAlternativeDerivationResponse2009, tintoLISASensitivitiesGravitational2010}. However, since the LISA constellation follows a heliocentric orbit, trailing Earth by approximately $20^\circ$, the spacecraft attain velocities on the order of $\beta = v/c \sim 10^{-4}$ in the solar system barycentric (SSB) frame \cite{martensTrajectoryDesignESA2021}. Velocity-dependent corrections to the link response have traditionally been neglected, as they are typically unobservable for most GW sources \cite{cornishLISAResponseFunction2003}. For the loudest MBHB signals, however, where precision is paramount, these corrections may exceed the noise floor and introduce biases in parameter estimation.

The state-of-the-art LISA response simulator \texttt{lisagwresponse}~\cite{bayleLISAGWResponse2023} computes the response for each link using the equation obtained under the assumption that spacecraft are stationary. However, their velocity is taken into account in the sense that the link vector connecting them is determined by interpolating the emitter's position back to time of emission instead of using its position at time of reception \cite{bayleSimulationDataAnalysis2019}. Velocity-dependent corrections to the response itself are not taken into account. We modify the response by including these corrections.

In Sec.~\ref{sec:derivation}, we derive expressions for the corrections to the LISA response function due to spacecraft velocity by determining both the timelike spacecraft geodesics as well as the null laser geodesic in the perturbed geometry. In Sec.~\ref{sec:results}, we evaluate the corresponding impact on the SNR for MBHBs. Concluding remarks are given in Sec.~\ref{sec:discussion}. App.~\ref{app:motion} provides a detailed exposition of the corrections to the spacecraft's geodesic and App.~\ref{app:TTF} presents an alternative derivation of the velocity contributions using the time transfer function framework. App.~\ref{app:LF} includes the low-frequency approximation of the corrected response function we have obtained.

Throughout this work we use the $(-+++)$ metric signature. Four-dimensional quantities are denoted using Greek indices $\mu, \nu, \ldots$ or boldface symbols $\mathbf{x}$, while three-dimensional spatial quantities use Latin indices $i, j, \ldots$ or arrow notation $\vec{x}$. The dot product $\vec{x}\cdot\vec{y}$ of such spatial vectors is computed using the flat background metric $\delta_{ij}$. We denote the spacecraft index $I=1,2,3$ using capital Latin indices.

\section{\label{sec:derivation} Derivation of Modified LISA Response}

In a space-based GW detector, the relevant observable is the frequency shift of the light ray exchanged between two test masses in geodesic motion. These frequency shifts occur over a wide range of frequencies and are not only generated by GWs but, for instance, also by Doppler contributions arising from the relative motion of the test masses.

To derive an expression for the frequency shifts, we consider an emitting spacecraft (or rather freely-falling test mass) at position $\vec{X}_e$ that sends a laser at coordinate time $t_e$ to a receiving spacecraft at $\vec{X}_r$, which receives the signal at $t_r$. Either the emission time $t_e$ and position $\vec{X}_e(t_e)$, or the reception time $t_r$ and position $\vec{X}_r(t_r)$ can be treated as being known. To match the conventions used in \texttt{lisagwresponse}, we work from the perspective of the receiver and we treat the reception time $t_r$ as given and use it to parameterize the system in time. The emission time $t_e$ and derived quantities are then determined perturbatively.

Now, we introduce a GW $h_{\mu\nu}$, which we assume to be a plane wave propagating in the $\hat{k}$-direction. Care must be taken in specifying the gauge and reference frame when treating GWs together with spacecraft velocities. The GW is most conveniently described in the transverse-traceless (TT) gauge. Within the TT gauge, there is still freedom in the choice of reference frame~\cite{schutzFirstCourseGeneral2022}. For LISA simulation purposes, we adopt the ICRS frame, in which the solar system's barycenter (SSB) is at rest at the origin \cite{ariasExtragalacticReferenceSystem1995, lisaddpcconventionsworkinggroupLISARosettaStone2025}. In this frame, the LISA spacecraft move at speeds of order $\beta=v/c\sim 10^{-4}$ along their one-year orbits. Because the GW field is defined entirely in this frame, the response we derive depends only on the spacecraft motion relative to the SSB: any motion of the source with respect to the SSB will be encoded in the waveform $h_{\mu\nu}^\mathrm{SSB}(t)$ \cite{torres-orjuelaDetectingBeamingEffect2019, torres-orjuelaMovingGravitationalWave2023}. 

The ICRS frame is described by the orthonormal basis $(\hat{x},\hat{y},\hat{z})$, with $\hat{x}$ and $\hat{y}$ spanning the equatorial plane \cite{ariasExtragalacticReferenceSystem1995, lisaddpcconventionsworkinggroupLISARosettaStone2025}. Suppose the GW source is located at the right ascension (equatorial longitude) $\alpha$ and  declination (equatorial latitude) $\delta$. We then define an orthonormal basis $(\hat{u},\hat{v},\hat{k})$ naturally adapted to the GW, related to the ICRS frame by
\begin{equation}
    \begin{split}
    \hat{u} &= (\sin \alpha, -\cos \alpha, 0),\\
    \hat{v} &= (-\sin \delta \cos \alpha, -\sin \delta \sin \alpha, \cos \delta),\\
    \hat{k} &= (-\cos\delta \cos \alpha, -\cos \delta \sin \alpha, -\sin \delta).
    \end{split}
    \label{eq:orthonormal}
\end{equation}

To the order $\mathcal O(\beta h)$, where $h$ is the typical GW strain amplitude, the GW will impact both the geodesic motion of the test masses and the light propagation. In Sec.~\ref{sec:trajectory}, we model the impact of the GW on the motion of the test masses. In Sec.~\ref{sec:light}, we consider the impact of the GW on the light propagation and, subsequently, we combine these two results in Sec.~\ref{sec:combine}.

\subsection{Modeling of the test-mass trajectory}\label{sec:trajectory}
In the Solar System, the spacetime metric which accounts for both the GW and the unperturbed gravitational field can be written, to linear order as $g_{\mu\nu}=\eta_{\mu\nu}+h_{\mu\nu}^\mathrm{SS}+h_{\mu\nu}$, where $h_{\mu\nu}^\mathrm{SS}$ is the metric perturbation corresponding to the Solar System and $h_{\mu\nu}$ is the GW. The two test masses follow a geodesic $\mathbf{X}$. Using the TT gauge and expressing the geodesic equation in coordinate time $t$ to first order in the GW strain, one gets the additional acceleration induced by the GW (see e.g. Eq. (2.2.61) from \cite{brumberg:1991uq} or Appendix B from \cite{torres-orjuelaDetectingBeamingEffect2019}):
\begin{equation}
    \left[ X^{\dprime\, i}\right]_\mathrm{GW} = -  h_{lm}^\prime \delta^{il}  X^{\prime\, m} \, ,
\end{equation}
where higher orders terms in the velocity (i.e. terms relatively smaller by an additional factor $\mathcal O(\beta)$) have been neglected. In these expressions, a prime denotes a derivative with respect to the coordinate time $t$. The test mass motion is therefore solution to 
\begin{equation}\label{eq:acc_pert}
 X^{\dprime\,i} = a^i -  h_{lm}^\prime \delta^{il}  X^{\prime\, m}\, ,
\end{equation}
where $a^i$ is the unperturbed acceleration whose leading contribution is the Newtonian interaction from the Sun, see \cite{martensTrajectoryDesignESA2021} for a detailed discussion. It is convenient to solve this equation of motion perturbatively in the GW strain $h_{ij}$. The zeroth-order solution $X^i_{(0)}$ is solution to the usual Solar System equations of motion \cite{martensTrajectoryDesignESA2021}. In the context of LISA, these solutions can be provided by the \texttt{LISAOrbits} Python package~\cite{bayleLISAOrbits2022}. The first-order solution $X^i_{(1)}$ is then a solution to
\begin{equation}
    X^{\dprime\,i}_{(1)} = \left.\frac{\partial a^i}{\partial X^j}\right|_{X=X_{(0)}} X^j_{(1)} - h_{ij}^\prime X_{(0)}^{\prime\,j} \, .
    \label{eq:EOM_SC}
\end{equation}
This equation depends on two different time scales, the orbital time scales characterized by $\fo=1/\SI{1}{yr}\approx \SI{3.2e-8}{Hz}$ and the GW frequency $\fg$ which is between $10^{-4}$ and $\SI{e-1}{Hz}$ in the case of LISA.

In appendix~\ref{app:motion}, we demonstrate numerically and by an analytical argument that in the case where the two time scales are well separated (i.e. $\fo \ll \fg$), the solution to Eq.~\eqref{eq:EOM_SC} can be written as a sum of two terms: one large out-of-band component (which oscillates at $f\sim \fo$) and an in-band component which is a solution to 
\begin{equation}\label{eq:ddotxi}
    X^{\dprime\,i}_{(1)} = -  h_{lm}^\prime \delta^{il}  X^{\prime\, m}_{(0)}\, .
\end{equation}
In doing this approximation, we neglect terms in-band that are relatively smaller by a factor $\left(\fo/\fg\right)^2$ as shown in Appendix~\ref{app:motion}.

The last equation can be solved numerically, but an integration by parts leads to an analytical solution that can also be used. Indeed, integrating Eq.~(\ref{eq:ddotxi}) leads to
\begin{equation}
X^{\prime\,i}_{(1)}(t) = -  h_{lm} \delta^{il}  X^{\prime\, m}(t) - \int \mathrm{d}\tau \:  h_{lm} \delta^{il}  X_{(0)}^{\dprime\,m}(\tau) \, .
\end{equation}
The first term of this equation is of the order of $\mathcal O(hv)$, where $h$ is the GW amplitude and $v$ the typical spacecraft velocity, while the order of magnitude of the second term is $\mathcal O(hv \fo /\fg)$, i.e. a factor $\fo/\fg\sim 10^{-8}-10^{-4}$ smaller than the first term for GW sources in the LISA band. Neglecting this second term leads to the following solution for the in-band first-order spacecraft velocity 
\begin{equation}\label{eq:GW_velocity}
 X^{\prime\,i}_{(1)} = -  h_{lm} \delta^{il}  X_{(0)}^{\prime\,m} \, .
\end{equation}
The solution in terms of position can be obtained by further integrating this last equation but will not be needed as we shall see in Sec.~\ref{sec:combine} and App.~\ref{app:pointing}.

To summarize, the impact of the GW on the spacecraft velocity can be approximated by Eq.~(\ref{eq:GW_velocity}) if: (i) the GW frequency is way larger than the orbital frequency, (ii) one is interested only in the in-band component of the velocity (a large out-of-band component at $f\sim \fo$ is neglected).

\subsection{Modeling of light-propagation}\label{sec:light}

\begin{figure}[b]
    \centering
    \begin{adjustbox}{width=0.7\linewidth}
    \begin{tikzpicture}
        \draw[->, thick] (-0.1,0) -- (5,0) node[anchor=west]{};
        \draw[->, thick] (0,-4.5) -- (0,0.3) node[anchor=south]{$ct$};

        \draw[dashed] (1.5,-4.5) -- (1.5,0);
        \draw[dashed] (4.5,-4.5) -- (4.5,0);

        \draw[thick] (0,0) -- (-0.1,0) node[left]{$ct_r$};
        \draw[thick] (0,-3) -- (-0.1,-3);
        \draw[dashed] (0,-3) -- (5,-3);
        \draw[thick] (0,-3.5) -- (-0.1,-3.5) node[left]{$ct_e^{(0)}$};
        \draw[dashed] (0,-3.5) -- (5,-3.5);
        \draw[thick] (0,-3.85) -- (-0.1,-3.85) node[left]{$ct_e$};
        \draw[dashed] (0,-3.85) -- (5,-3.85);

        \node[circle, fill=black,inner sep=1pt, label={above}:{$\vec{X}_r(t_r)$}] (Xr) at (4.5,0) {};
        \node[circle, fill=black,inner sep=1pt, label={above}:{$\vec{X}_e(t_r)$}] (Xer) at (1.5,0) {};
        \draw[very thick, red] (0.86,-4.5) -- (Xer);
        \draw[very thick, red] (4.3,-4.5) -- (Xr);
        
        \node[circle, fill=black,inner sep=1pt] (Xee0) at (1,-3.5) {};
        \node[circle, fill=black,inner sep=1pt] (Xee1) at (0.95,-3.85) {};
        
        \draw[thick, blue] (Xee0) -- (Xr);
        \draw[thick, blue, decorate,decoration={snake,amplitude=.3mm}] (Xee1) -- (Xr);

        \draw[decoration={brace,raise=2pt},decorate] (0,0) -- node[right=3pt] {$D_{(0)}$} (0,-3);
        \draw[decoration={brace,raise=2pt},decorate] (0,-3) -- node[right=3pt] {$\Delta D$} (0,-3.5);
        \draw[decoration={brace,raise=2pt},decorate] (0,-3.5) -- node[right=4pt] {$c\delta t$} (0,-3.85);
    \end{tikzpicture}
    \end{adjustbox}
    \caption{Spacetime diagram of laser propagation from the emitter at position $\vec{X}_e(t_e)$ to the receiver at $\vec{X}_r(t_r)$. Spacecraft worldlines are depicted in red. The laser's null rays are shown in blue: the straight line represents the trajectory in flat spacetime and the wavy line the trajectory in the perturbed geometry.}
    \label{fig:spacetimediagram}
\end{figure}

In this subsection, we model the light propagation between the two spacecraft, assuming we know their trajectories and velocities, i.e. we focus on the impact of the GW on the null geodesic connecting the two spacecraft, including corrections linear in the spacecraft velocity. We will derive this frequency shift by making use of Killing vectors, a method originally developed in the case of stationary test masses in~\cite{estabrookResponseDopplerSpacecraft1975, burkeLargeScaleRandomGravitational1975, wahlquistDopplerResponseGravitational1987} and later also adopted in, e.g.,~\cite{cornishAlternativeDerivationResponse2009, tintoLISASensitivitiesGravitational2010}. This method does not require any assumptions about the GW's wavelength relative to the detector's arm length.

As a plane wave, the GW is parameterized by the retarded time $\xi = ct - \hat{k}\cdot\vec{x}$ describing its wavefronts. We therefore introduce null coordinates $\xi = ct - k$ and $\eta = ct + k$, in which the metric in the polarization basis Eq.~\eqref{eq:orthonormal} takes the form
\begin{equation}
\begin{split}
    ds^2 =& - \: d\xi\, d\eta + 2\, h_\times(\xi) \: du\, dv\\
     &+ [1+h_+(\xi)] \: du^2 + [1-h_+(\xi)] \: dv^2.
\end{split}
\end{equation}
In this form, it is evident that the geometry admits three Killing vectors: $\partial_u$, $\partial_v$, and $\partial_\eta = \partial_{ct} + \partial_k$. These symmetries allow us to find the null geodesic connecting the spacecraft from first-order differential equations~\cite{cornishAlternativeDerivationResponse2009}, instead of solving the second-order geodesic equations explicitly~\cite{finnResponseInterferometricGravitational2009}. We denote the laser’s null geodesic connecting the emitter and receiver by $\boldsymbol{\sigma}(\lambda)$, parametrized by affine parameter $\lambda$. The three Killing vectors give rise to three conserved constants of motion along $\boldsymbol{\sigma}$:
\begin{align}
    (1+h_{+})\dot{\sigma}^u + h_{\times}\dot{\sigma}^v &= \alpha_1, \label{eq:alpha1}\\
    (1-h_{+})\dot{\sigma}^v + h_{\times}\dot{\sigma}^u &= \alpha_2, \label{eq:alpha2}\\
    -\tfrac{1}{2} \dot{\sigma}^\xi &= \alpha_3, \label{eq:alpha3}
\end{align}
where $\dot{\boldsymbol{\sigma}}=d\boldsymbol{\sigma}/d\lambda$ is the geodesic's tangent vector. The fact that $\boldsymbol{\sigma}$ is null imposes the condition
\begin{equation}
\begin{split}
    -\dot{\sigma}^\xi \dot{\sigma}^\eta + (1+h_{+})(\dot{\sigma}^u)^2 + (1-h_{+})(\dot{\sigma}^v)^2 & \\ 
    + 2 h_{\times} \dot{\sigma}^u \dot{\sigma}^v &= 0,
    \end{split}
    \label{eq:null}
\end{equation}
which, using Eqs.~\eqref{eq:alpha1}--\eqref{eq:alpha3}, can be rewritten as
\begin{equation}
    2\alpha_3 \dot{\sigma}^\eta + \alpha_1 \dot{\sigma}^u + \alpha_2 \dot{\sigma}^v = 0.
    \label{eq:nullalt}
\end{equation}
Finally, the geodesic must intersect the spacecraft trajectories at its endpoints, i.e., $\vec{\sigma}(\lambda_e)=\vec{X}_e(t_e)$ and $\vec{\sigma}(\lambda_r)=\vec{X}_r(t_r)$, which provides a complete system of equations for $\boldsymbol{\sigma}(\lambda)$.

\subsubsection{Geodesic in Minkowski Spacetime}\label{subsection: zeroth order}

First, we determine the laser beam’s geodesic in the Minkowski background. At zeroth-order in the metric perturbation $h$, Eqs.~\eqref{eq:alpha1}–\eqref{eq:alpha3} reduce to
\begin{align}
\dot{\sigma}^u_{(0)} &= \alpha_1^{(0)}, \label{eq:alpha1O0}\\
\dot{\sigma}^v_{(0)} &= \alpha_2^{(0)}, \label{eq:alpha2O0}\\
\dot{\sigma}^\xi_{(0)} &= -2\alpha_3^{(0)}, \label{eq:alpha3O0}
\end{align}
where the superscript $(i)$ indicates the order in the perturbative expansion in $h$. The boundary terms are determined by the intersections of this null geodesic with the spacecraft timelike geodesics, i.e. $\vec{\sigma}(\lambda_e)=\vec{X}_e(t_e^{(0)})$ and $\vec{\sigma}(\lambda_r)=\vec{X}_r(t_r)$. The reception time $t_r$ is fixed, irrespective of approximations, and the emission time $t_e^{(0)}$ is unknown. Let $L_{(0)}$ denote the Euclidean distance between $\vec{X}_r(t_r)$ and $\vec{X}_e(t_e^{(0)})$. Eqs.~\eqref{eq:alpha1O0}--\eqref{eq:alpha3O0} together with the null condition Eq.~\eqref{eq:nullalt} are straightforward to solve.  The resulting geodesics, in $(ct,u,v,k)$ coordinates, are straight lines:
\begin{align}
    \dot{\sigma}^t_{(0)}(\lambda) &= \frac{L_{(0)}}{\lambda_r-\lambda_e},\\
    \dot{\sigma}^j_{(0)}(\lambda) &= \frac{L_{(0)}\hat{n}^j_{(0)}}{\lambda_r-\lambda_e},
\end{align}
where $\hat{n}_{(0)}$ is the unit vector pointing from $\vec{X}_e(t_e^{(0)})$ to $\vec{X}_r(t_r)$, i.e.
\begin{equation}\label{eq:pointing_vec}
    \hat{n}_{(0)} = \frac{\vec{X}_r(t_r) -\vec{X}_e(t_e^{(0)}) }{\left|\vec{X}_r(t_r) -\vec{X}_e(t_e^{(0)})\right|}\, . 
\end{equation}

So far, the zeroth-order equations in this section hold irrespective of spacecraft velocity. To make the velocity dependence explicit, let us now expand to first-order in $\beta$. In our perturbative set-up, we only have direct access to a snapshot of the constellation at $t_r$, so we only know the instantaneous separation $D_{(0)}$ between $\vec{X}_e(t_r)$ and $\vec{X}_r(t_r)$ and the corresponding unit vector $\hat{m}_{(0)}$ connecting these points, as illustrated in Fig.~\ref{fig:spacetimediagram}. To determine the correct separation $L_{(0)}=D_{(0)}+\Delta D$, we expand up to linear order in $\beta_e$, which we regard as constant during the short \SI{8}{\second} light travel time. Note that for $\beta_e\ll 1$, we can approximate the change of path length $\Delta D$ by projecting the displacement $D_{(0)}\vec{\beta}_e$ of the emitter during the light travel time onto the line of sight $\hat{n}_{(0)}$. Expanding this up to linear order, we find
\begin{equation}
    \Delta D = D_{(0)} \: \vec{\beta}_e\cdot \hat{n}_{(0)} +\mathcal{O}(\beta^2) =D_{(0)}\:\vec{\beta}_e\cdot\hat{m}_{(0)}+\mathcal{O}(\beta^2).
    \label{eq:Lapprox} 
\end{equation}
The emitter's velocity also changes the aiming direction through a ``point-ahead'' correction. To determine $\hat{n}_{(0)}$, we note that the emitter's position can be expanded as $\vec{X}_e(t_e^{(0)})=\vec{X}_e(t_r)-L_{(0)}\vec{\beta}_e+\mathcal{O}(\beta^2)$. Substituting this in the expression for $\hat{n}_{(0)}$ and expanding this, yields
\begin{equation}
    \hat{n}_{(0)}=\hat{m}_{(0)}+\beta_e-(\vec{\beta}_e\cdot\hat{m}_{(0)})\: \hat{m}_{(0)}+\mathcal{O}(\beta^2),
    \label{eq:napprox}
\end{equation}
where the correction is the component of the velocity perpendicular to the line of sight. This fully specifies the laser's geodesic in the unperturbed geometry to linear order in the velocity.

\subsubsection{Geodesic in Perturbed Geometry}
\label{subsection: first order}
When a GW $h_{\mu\nu}$ passes through the constellation, the laser's null trajectory is lenzed, imparting a time delay (or contraction) $\delta t$. To determine this perturbed geodesic and, consequently, the time delay, we solve the system of equations linear in $h$:
\begin{align}
    \dot{\sigma}_{(1)}^u + h_{+}\dot{\sigma}_{(0)}^u + h_{\times} \dot{\sigma}_{(0)}^v &=\alpha_1^{(1)}, \label{eq:alpha1O1}\\
    \dot{\sigma}_{(1)}^v - h_{+}\dot{\sigma}_{(0)}^v + h_{\times} \dot{\sigma}_{(0)}^v &=  \alpha_2^{(1)}, \label{eq:alpha2O1}\\
    -\tfrac{1}{2}\dot{\sigma}_{(1)}^\xi &= \alpha_3^{(1)}. \label{eq:alpha3O1}
\end{align}
We assume the spacecraft trajectories $\vec{X}_I$ to be known perfectly (see Sec.~\ref{sec:trajectory} for their expansion in $\beta$ and $h$), and hence we can specify the boundary conditions. Due to the time delay or contraction imparted by the GW during propagation, the null geodesic intersects the timelike geodesic at different points, compared to the unperturbed case (see Fig.~\ref{fig:spacetimediagram}). Since we assume the receiver position to be given, we have the boundary condition $\vec{\sigma}_{(1)}(\lambda_r)=0$. The signal should have been emitted at emission time $t_e=t_e^{(0)}-\delta t+\mathcal{O}(h^2)$, which corresponds to the emission position $\vec{X}_e(t_e)=\vec{X}_e(t_e^{(0)})-c\delta t\:\vec{\beta}_e+\mathcal{O}(\beta^2 h)$. This gives rise to the boundary condition $\vec{\sigma}_{(1)}(\lambda_e)=-c\delta t\:\vec{\beta}_e$. 

First, we solve for the constant $\alpha_3^{(1)}$ by integrating Eq.~\eqref{eq:alpha3O1} and find
\begin{equation}
    \alpha_3^{(1)}=-\frac{c\delta t(1-\hat{k}\cdot\vec{\beta}_e)}{2(\lambda_r-\lambda_e)}
\end{equation}
Next, we integrate Eqs.~\eqref{eq:alpha1O1}–\eqref{eq:alpha2O1} with respect to $\lambda$. Since the GW is more naturally parameterized by its wavefronts $\xi$, we change variables in the integrals of $h$, noting that to zeroth-order
\begin{equation*}
    \frac{d\lambda}{d\xi} =\frac{\lambda_r-\lambda_e}{L_{(0)}(1-\hat{k}\cdot\hat{n}_{(0)})}
    +\mathcal{O}(h).
\end{equation*}
This yields the corrections to the constant of motion:
\begin{align}
    \alpha_1^{(1)}=\frac{\delta^{ul}\hat{n}_{(0)}^m}{(\lambda_r-\lambda_e)(1-\hat{k}\cdot\hat{n}_{(0)})}H_{lm}+\frac{c\delta t \: \hat{u}\cdot\vec{\beta}_e}{\lambda_r-\lambda_e},\\
    \alpha_2^{(1)}=\frac{\delta^{vl}\hat{n}_{(0)}^m}{(\lambda_r-\lambda_e)(1-\hat{k}\cdot\hat{n}_{(0)})}H_{lm}+\frac{c\delta t \: \hat{v}\cdot\vec{\beta}_e}{\lambda_r-\lambda_e},
\end{align}
where we abbreviate the integral over the unperturbed geodesic as $H_{ij}\equiv \int_{\xi_e}^{\xi_r}\mathrm{d}\xi \:h_{ij}(\xi)\lvert_{\boldsymbol{\sigma}_{(0)}}$. The time delay $\delta t$ follows from integrating the null condition \eqref{eq:nullalt}. Up to linear order in both $h$ and $\beta$, we find
\begin{equation}
    c\delta t = (1+\vec{\beta}_e\cdot\hat{n}_{(0)})\frac{1}{2}\frac{\hat{n}^l_{(0)}\hat{n}^m_{(0)}}{1-\hat{k}\cdot \hat{n}_{(0)}}H_{lm}+\mathcal{O}(\beta^2 h)
    \label{eq:deltat}
\end{equation}
Here, we recognize the delay obtained for stationary spacecraft~\cite{cornishLISAResponseFunction2003, cornishAlternativeDerivationResponse2009} plus an additional contribution of the velocity projected along the line of sight. As a validation, this result has also been obtained using the time transfer formalism in App.~\ref{app:TTF}, see Eq.~\eqref{eq:trG1}. The constants of motion and $\delta t$ fully determine the first-order corrections to the null geodesic:
\begin{align*}
    \dot{\sigma}^t_{(1)}(\lambda)=&\frac{(1-\frac{1}{2}\hat{k}\cdot\hat{n}_{(0)})\:\hat{n}_{(0)}^l\hat{n}_{(0)}^m}{(\lambda_r-\lambda_e)(1-\hat{k}\cdot\hat{n}_{(0)})^2}H_{lm} \label{eq:DsigmatO1}\\
    &+\frac{1}{2}\frac{\hat{n}_{(0)}\cdot\vec{\beta}_e \:\hat{n}_{(0)}^l\hat{n}_{(0)}^m}{(\lambda_r-\lambda_e)(1-\hat{k}\cdot\hat{n}_{(0)})}H_{lm}\\
    &-\frac{1}{2}\frac{L_{(0)}\hat{n}_{(0)}^l\hat{n}_{(0)}^m}{(\lambda_r-\lambda_e)(1-\hat{k}\cdot\hat{n}_{(0)})}h_{lm}, \\
    \dot{\sigma}^j_{(1)}(\lambda)=&\left(\delta^{jm}+\frac{1}{2}\frac{\hat{n}_{(0)}^m\delta^{jp}\hat{k}_p}{1-\hat{k}\cdot\hat{n}_{(0)}}\right)\frac{\hat{n}_{(0)}^lH_{lm}}{(\lambda_r-\lambda_e)(1-\hat{k}\cdot\hat{n}_{(0)})}\\
    &+\frac{1}{2}\frac{\vec{\beta}_e^j \:\hat{n}_{(0)}^l\hat{n}_{(0)}^m}{(\lambda_r-\lambda_e)(1-\hat{k}\cdot\hat{n}_{(0)})}H_{lm}\\
    &-\left(\delta^{jm}\hat{n}_{(0)}^l+\frac{1}{2}\frac{\delta^{jp}\hat{k}_p\hat{n}_{(0)}^l\hat{n}_{(0)}^m}{1-\hat{k}\cdot\hat{n}_{(0)}}\right)\frac{L_{(0)}h_{lm}}{\lambda_r-\lambda_e}.
\end{align*}
Note that the terms proportional to $H_{lm}$ are independent of $\lambda$ and are therefore constant along the geodesic, while the terms proportional to $h_{lm}$ change along the geodesic. At this stage, we still have freedom in the choice of parameterization of $\lambda$. We select the parameterization such that
\begin{align*}
    \lambda_r-\lambda_e =& L_{(0)}+(1-\frac{1}{2}\hat{k}\cdot\hat{n}_{(0)})\frac{\hat{n}_{(0)}^l\hat{n}_{(0)}^mH_{lm}}{(1-\hat{k}\cdot\hat{n}_{(0)})^2}\\
    &+\frac{1}{2}\frac{\hat{n}_{(0)}\cdot\vec{\beta}_e \:\hat{n}_{(0)}^l\hat{n}_{(0)}^m}{(\lambda_r-\lambda_e)(1-\hat{k}\cdot\hat{n}_{(0)})}H_{lm}+ \mathcal{O}(\beta h).
\end{align*}
This choice simplifies the time component of $\dot{\boldsymbol{\sigma}}$ significantly. It is consistent with the derivation in \cite{finnResponseInterferometricGravitational2009}, where the geodesic equation is solved directly. Moreover, it makes the derivation of the response function more tractable. Importantly, the final expression for the response is independent of the parameterization and can equally well be obtained from the general expression for $\dot{\boldsymbol{\sigma}}$ given above. With this choice, the geodesic reduces to
\begin{align}
    \dot{\sigma}^t(\lambda) =& 1-\frac{1}{2}\frac{\hat{n}_{(0)}^l\hat{n}_{(0)}^m}{1-\hat{k}\cdot\hat{n}_{(0)}}h_{lm},\\
    \dot{\sigma}^j(\lambda) =& \hat{n}_{(0)}^j + \delta \hat{n}^j \nonumber\\ 
    &-\left(\hat{n}^l_{(0)}\delta^{jm}+\frac{1}{2}\frac{\hat{n}_{(0)}^l\hat{n}_{(0)}^m}{1-\hat{k}\cdot\hat{n}_{(0)}}\delta^{jp}\hat{k}_p\right)h_{lm}.
\end{align}
Here, we have introduced the vector
\begin{align}
    \delta\hat{n}^j=&(\delta^{jp}-\hat{n}_{(0)}^j\hat{n}^p_{(0)})\left(\delta^m_p+\frac{1}{2}\frac{\hat{n}_{(0)}^m\hat{k}_p}{1-\hat{k}\cdot{n}_{(0)}}\right)\times \nonumber\\
    &\frac{\hat{n}_{(0)}^lH_{lm}}{L_{(0)}(1-\hat{k}\cdot\hat{n}_{(0)})}
    +\frac{c\delta t}{L_{(0)}}(\beta_e^j-\vec{\beta}_e\cdot\hat{n}_{(0)}\:\hat{n}_{(0)}^j).
    \label{eq:n1}
\end{align}
This vector $\delta \hat{n}$ encodes a correction to the line-of-sight, and contains two contributions: a GW-induced angular deflection of the laser’s path in the first term~\cite{finnResponseInterferometricGravitational2009}, and a point-ahead correction proportional to $\beta_e$ in the second term. Together with $\hat{n}_{(0)}$, the corrected line-of-sight composes a null vector $\mathbf{n}=(1,\hat{n}_{(0)}+\delta\hat{n})$ in the unperturbed geometry.

The response function $y=(\nu_r-\nu_e)/\nu_e$ is a fractional frequency shift, where $\nu_r$ is the laser's frequency as observed by the receiving spacecraft and $\nu_e$ the frequency observed by the emitter. For a timelike observer with four-velocity $\mathbf{U}$, the observed frequency is $\nu=-g_{\mu\nu}U^\mu P^\nu$, where $\mathbf{P}=\tfrac{\nu_0}{c}\dot{\boldsymbol{\sigma}}$ is the photon's propagation vector. Here, $\nu_0$ is the frequency as observed by a static observer, located at infinity. Each spacecraft observes a frequency:
\begin{equation}
    \nu_I=\gamma\nu_0(\dot{\sigma}^t-\delta_{ij}\beta^i_I\dot{\sigma}^j-h_{ij}\beta^i_I\dot{\sigma}^j_{(0)})+\mathcal{O}(h^2).
\end{equation}
We can now determine the GW response function by comparing $\nu_r$ and $\nu_e$. Here, we can neglect $\mathcal{O}(\beta^2)$ Lorentz factors. The final result, i.e. the GW contribution to the one-way frequency shift, which is expanded up to linear order in both $h$ and $\beta$, is
\begin{widetext}

\begin{equation}
    \begin{split}
        y_\mathrm{GW}^{\mathrm{propag}}=&-\frac{1}{2}\frac{\hat{n}_{(0)}^l\hat{n}_{(0)}^m}{1-\hat{k}\cdot\hat{n}_{(0)}}[h_{lm}(\xi_r)-h_{lm}(\xi_e)]
        +\frac{1}{2}\frac{\hat{n}_{(0)}^l\hat{n}_{(0)}^m}{1-\hat{k}\cdot\hat{n}_{(0)}}[\vec{\beta}_r\cdot\hat{k}\:h_{lm}(\xi_r)-\vec{\beta}_e\cdot\hat{k}\:h_{lm}(\xi_e)]\\
        &-\frac{1}{2}\frac{\hat{n}_{(0)}^l\hat{n}_{(0)}^m}{1-\hat{k}\cdot\hat{n}_{(0)}}\hat{n}_{(0)}\cdot\vec{\beta}_e[h_{lm}(\xi_r)-h_{lm}(\xi_e)]
        -\frac{1}{2}\frac{\hat{n}_{(0)}^l\hat{n}_{(0)}^m}{1-\hat{k}\cdot\hat{n}_{(0)}}\hat{n}_{(0)}\cdot(\vec{\beta}_r-\vec{\beta}_e) h_{lm}(\xi_e)\\
        &-\delta_{ij}(\beta^i_r-\beta_e^i)\delta\hat{n}^j+\mathcal{O}(\beta^2 h). 
    \end{split}
    \label{eq:response}
\end{equation}

\end{widetext}
Note this expression can also be obtained by taking the derivative of the time delay in Eq.~\eqref{eq:deltat} with respect to the reception time $t_r$ (see App.~\ref{app:TTF} for more details). Furthermore, at zeroth-order in $h$ we recover the laser's Doppler shift 
\begin{equation}\label{eq:Doppler0}
y_{(0)}=-\delta_{ij}(\beta_r^i-\beta_e^i)\hat{n}_{(0)}^j\, , 
\end{equation}
for the response function. 

The first term in Eq.~\eqref{eq:response} corresponds to the standard stationary test-mass GW response. The additional terms arise from the spacecraft velocities. We refer to the second term as the ``localized redshift'' correction (following the terminology of \cite{rakhmanovResponseTestMasses2005}). This terminology is motivated by the fact that the first two terms in Eq.~\eqref{eq:response}, taken together, are proportional to the contraction of the photon’s local tangent vector with the observer’s four-velocity, and thus effectively measure the redshift difference between emission and reception. The third and fourth terms can be interpreted as ``point-ahead'' corrections, since they account for the fact that the emitter’s position is not known directly. Specifically, the third term reflects the increased light-travel time due to the emitter’s motion, as indicated by the prefactor in Eq.~\eqref{eq:deltat}, while the fourth arises from the instant at which the wavefront $\xi_e$ intersects the emitter, as is more evident in App.~\ref{app:TTF}. The final term has a different character: it represents a modulation of the laser’s Doppler shift caused by the GW lensing the laser trajectory. This accumulated effect along the geodesic alters the apparent angle of arrival and, thus, the line-of-sight~\cite{finnResponseInterferometricGravitational2009}. Note that the $\beta$-dependent contribution in $\delta\hat{n}$ enters the response only at order $\beta^2 h$ and can therefore be neglected.

The perceived GW response is dependent on the sky-localization, depending on the orientation of the link with respect to the GW's propagation direction. Most terms follow the familiar antenna pattern~\cite{cornishLISAResponseFunction2003, bayleSimulationDataAnalysis2019}:
\begin{align}
    h_{lm}\hat{n}^l_{(0)}\hat{n}^m_{(0)} =& h_+ \left[(\hat{n}_{(0)}\cdot\hat{u})^2-(\hat{n}_{(0)}\cdot\hat{v})^2\right] \nonumber\\
    &+2h_\times (\hat{n}_{(0)}\cdot\hat{u})(\hat{n}_{(0)}\cdot\hat{v}).
\label{eq:antenna_nn}
\end{align}
The modulation of the Doppler term, on the other hand, contains a contraction of the accumulated GW $H_{ij}$ with the link- and velocity-vectors, which imposes a different antenna pattern. These contractions are of the form
\begin{align}
    H_{lm}\beta^l\hat{n}^m_{(0)} =& H_+ \left[(\vec{\beta}\cdot\hat{u})(\hat{n}_{(0)}\cdot\hat{u})-(\vec{\beta}\cdot\hat{v})(\hat{n}_{(0)}\cdot\hat{v})\right] \nonumber\\
    &+H_\times \left[(\vec{\beta}\cdot\hat{u})(\hat{n}_{(0)}\cdot\hat{v})+(\vec{\beta}\cdot\hat{v})(\hat{n}_{(0)}\cdot\hat{u})\right].
\label{eq:antenna_bn}
\end{align}

In summary, by solving the perturbed null geodesic to first-order in $h$ and $\beta$, we obtain a closed-form expression for the GW-induced frequency shift due to the laser propagation that naturally incorporates Doppler shifts of the GW, point-ahead corrections and modulation of the laser's Doppler shift. This extends the standard stationary result to moving spacecraft.

\subsection{Summary and final result}\label{sec:combine}

Section \myhyperref{sec:trajectory} derives the first-order $\mathcal{O}(\beta h)$ correction to spacecraft trajectories and velocities $\vec{\beta}_{r, (1)} = \tfrac{1}{c}\vec{X}^{\prime}_{r,(1)}$ and $\vec{\beta}_{e, (1)} = \tfrac{1}{c}\vec{X}^{\prime}_{e,(1)}$, perturbing the end-points of the light propagation between the reception and emission events. On the other hand, Section \ref{sec:light} provides the $\mathcal{O}(\beta h)$ correction to the light propagation between two known a priori positions of the spacecraft at reception time. Say, in other words, Section \ref{sec:light} provides an expression of the light-travel time and frequency shift as a function of the (full) spacecraft trajectories. In this section, we will combine both results and introduce the spacecraft trajectory decomposition $\vec X = \vec{X}^{(0)} + \vec{X}^{(1)}$, into the expression of the frequency shift.

The corrections to the spacecraft trajectories are of order $\mathcal{O}(\beta h)$, as follows from Eq.~\eqref{eq:GW_velocity}, which can easily be propagated to the end points throughout the derivation from Section \ref{sec:light}. What appears in the expression of the frequency shift given in Eqs.~(\ref{eq:response}) and (\ref{eq:Doppler0}) is the spacecraft velocities and the link vector. The correction on the spacecraft trajectories derived in Sec. \ref{sec:trajectory} leads to an additional correction $\hat{n}_{(1)}$ of order $\mathcal{O}(\beta h)$ to the link vector connecting the spacecraft. However, as shown in App.~\ref{app:pointing}, this correction $\hat{n}_{(1)}$ is suppressed by a factor $f_\mathrm{orb}/f_\mathrm{GW}$ and can be safely neglected in the corrected response. Likewise, propagating the corrections $\vec{\beta}^{(1)}_I$ to the spacecraft velocities throughout Eq.~(\ref{eq:response})  would lead to  next-order terms at $\mathcal{O}(\beta h^2)$, which are also negligible. Hence, the only term being impacted at $\mathcal{O}(\beta h)$ order by the GW acceleration of the spacecraft is the correction to the $\beta^i_I$ which appears in the out-of-band Doppler term from Eq.~(\ref{eq:Doppler0}).
This term now gets an additional, in-band $\mathcal{O}(\beta h)$ contribution:
\begin{align}
    y_\mathrm{GW}^{\mathrm{SC}} &= -\delta_{ij} \left[ \beta^i_{r,(1)}-\beta_{e,(1)}^i \right] \hat{n}^j_{(0)} 
    \nonumber\\
    &\approx \left[ \beta^i_{r,(0)} h_{ij}(\xi_r) - \beta^i_{e,(0)} h_{ij}(\xi_e) \right] \hat{n}^j_{(0)}\, .
    \label{eq:sc_acc_doppler}
\end{align}

Combining $y_\mathrm{GW}^{\mathrm{SC}}$ in Eq.~\eqref{eq:sc_acc_doppler} with $y_\mathrm{GW}^{\mathrm{propag}}$ from Eq.~\eqref{eq:response}, one gets the net response between spacecraft at $\mathcal{O}(\beta h)$ order:
\begin{equation}
    y_\mathrm{GW} = y_\mathrm{GW}^{\mathrm{SC}} + y_\mathrm{GW}^{\mathrm{propag}} +\mathcal{O}(\beta^2 h)\, .
\end{equation}
More explicitly, we find the full response function up to linear order in both  $h$ and $\beta$ to be:
\begin{widetext}
\begin{equation}
    \begin{split}
        y_\mathrm{GW}=&-\frac{1}{2}\frac{\hat{n}_{(0)}^l\hat{n}_{(0)}^m}{1-\hat{k}\cdot\hat{n}_{(0)}}[h_{lm}(\xi_r)-h_{lm}(\xi_e)]
        +\frac{1}{2}\frac{\hat{n}_{(0)}^l\hat{n}_{(0)}^m}{1-\hat{k}\cdot\hat{n}_{(0)}}[\vec{\beta}_{r}^{(0)}\cdot\hat{k}\:h_{lm}(\xi_r)-\vec{\beta}_{e}^{(0)}\cdot\hat{k}\:h_{lm}(\xi_e)]\\
        &-\frac{1}{2}\frac{\hat{n}_{(0)}^l\hat{n}_{(0)}^m}{1-\hat{k}\cdot\hat{n}_{(0)}}\hat{n}_{(0)}\cdot\vec{\beta}_{e}^{(0)}[h_{lm}(\xi_r)-h_{lm}(\xi_e)]
        -\frac{1}{2}\frac{\hat{n}_{(0)}^l\hat{n}_{(0)}^m}{1-\hat{k}\cdot\hat{n}_{(0)}}\hat{n}_{(0)}\cdot(\vec{\beta}_{r}^{(0)}-\vec{\beta}_{e}^{(0)}) h_{lm}(\xi_e)\\
        &-\delta_{ij}(\beta^i_{r,(0)}-\beta_{e,(0)}^i)\hat{n}^j_{(1)}
        +\left[ \beta^i_{r,(0)} h_{ij}(\xi_r) - \beta^i_{e,(0)} h_{ij}(\xi_e) \right] \hat{n}^j_{(0)}
        +\mathcal{O}(\beta^2 h). 
    \end{split}
    \label{eq:full_response}
\end{equation}
\end{widetext}

\section{\label{sec:results} Impact on observed SNR}

The velocity-dependent corrections to the LISA response function introduced in Eq.~\eqref{eq:full_response} are several orders of magnitude smaller (on the order of $\sim 10^{-4}$) than the leading terms. As a result, their effect on the total SNR is expected to be below unity for most sources in the LISA band. Therefore, we focus our analysis on MBHBs, where signal strengths are highest, and investigate whether including the velocity-dependent terms leads to a significant increase in total SNR.

To simulate such events, we use the time domain phenomenological model \texttt{IMRPhenomTHM}~\cite{estellesTimedomainPhenomenologicalModel2022, estellesNewTwistsCompact2022} to generate full time-domain waveforms, including the inspiral, merger and ringdown. The generated waveforms used here contain harmonic modes up to $\ell\le 5$. In our set-up, we adopt a 2-year observation window and fix the merger-time, defined as the peak strain of the $(2,2)$-mode following standard convention, at exactly 1 year. Any signal laying outside of this window is truncated, resulting in an inspiral with a maximum duration of approximately 1 year. This duration is sufficient to capture the full signal in the time domain for sources with total redshifted masses $M_\mathrm{tot}\gtrsim \SI{e6}{\mathrm{M}_\odot}$. This symmetric window around the merger is chosen to mitigate edge effects when applying a window function prior to Fourier transformation in later analysis \cite{inchauspeMeasuringGravitationalWave2025}.  The waveforms have been sampled at \SI{0.2}{\hertz}.

The evolution of the LISA constellation during the observation period is modeled using orbit files generated with the \texttt{LISAOrbits} Python package~\cite{bayleLISAOrbits2022, martensTrajectoryDesignESA2021}, using the numerically optimized ESA trailing orbit configuration, sampled approximately once per day (\SI{e-5}{\hertz}). The time axis of the waveform is aligned with the orbit files such that the merger for all simulated events occurs at the same orientation of the LISA constellation.

We compute the time-domain response using the \texttt{lisagwresponse} package~\cite{bayleLISAGWResponse2023}, both in its original implementation and in a modified version that incorporates the velocity-dependent corrections from Eq.~\eqref{eq:full_response}. The original code already accounts for spacecraft motion by interpolating the emitter’s position $\vec{X}_e(t_e)$ back to the emission time, using the light-travel time from the orbit files. Our custom implementation instead applies the perturbative corrections derived in Eqs.~\eqref{eq:Lapprox} and \eqref{eq:napprox}. The spacecraft velocities are obtained as analytical derivatives of the spline interpolating the orbits and are evaluated at reception time, e.g. $\vec{\beta}_e \equiv \vec{\beta}_e(t_r)$. We have verified that both approaches to calculating $\hat{n}_{(0)}$ give identical results for the leading-order response, so differences between the methods arise only from the subleading terms in Eq.~\eqref{eq:full_response}. The integral $H_{ij}$ is evaluated analytically using the spline interpolants of $h_{ij}(\xi_e)$ and $h_{ij}(\xi_r)$ that are already computed for the baseline response: their antiderivatives are taken, and the results are subtracted. In the end, both response computations yield time series $y_{IJ}(t)$ describing the response of the link $IJ$ of a (distant) emitting spacecraft $J$ to a (local) receiving spacecraft $I$  \cite{lisaddpcconventionsworkinggroupLISARosettaStone2025}.

The individual link responses $y_{IJ}(t)$ are then combined using \texttt{PyTDI}~\cite{staabPyTDI2022} to form virtual interferometry channels that suppress laser and spacecraft longitudinal jitter noise. Specifically, we compute the second-generation $A$, $E$, and $T$ time-delay interferometry (TDI) variables~\cite{vallisneriGeometricTimeDelay2005, tintoTimedelayInterferometry2020}, which suppresses laser-noise below the mission's requirement in a realistic scenario~\cite{lisasciencestudyteamLISAScienceRequirements2018}.

The time domain TDI variables are transformed to the frequency domain using a Planck-taper window centered on the merger time~\cite{mckechanTaperingWindowTimedomain2010}. The SNR $\rho_C$ for each TDI channel $C \in \{A, E, T\}$ is then computed as:
\begin{equation}
    \rho_C^2 = 4 \Re\int_{f_{\min}}^{f_{\max}} \frac{\tilde{d}_C(f)\tilde{d}_C^*(f)}{S_{CC}(f)}\: df,
\end{equation}
where $\tilde{d}_C(f)$ is the positive-frequency Fourier transform of the data in channel $C$, and $S_{CC}(f)$ is the one-sided noise power spectral density (PSD) for that channel. We use $f_{\min} = \SI{1e-4}{\hertz}$ and $f_{\max} = \SI{1e-1}{\hertz}$.

The analytical PSD model is based on the SciRDv1 ``science requirement model''~\cite{lisasciencestudyteamLISAScienceRequirements2018, groupLISADataChallenge2022}, which includes the dominant secondary noise sources: test mass acceleration noise and optical metrology system noise. In addition, we include an analytical galactic confusion noise model assuming a 2-year observation time~\cite{groupLISADataChallenge2022}. To address numerical instabilities in computing the SNR arising at the zero-response frequencies, the PSD is smoothed by averaging twelve PSD realizations sampled monthly over one full year of orbital motion. 

Finally, we calculate the total SNR $\rho$ by combining the three channels $C\in\{A,E,T\}$, which we assume to be uncorrelated, and summing them quadratically:
\begin{equation}
    \rho=\sqrt{\sum_{C\in\{A,E,T\}}\rho_C^2}.
\end{equation}

\begin{figure*}
    \centering
    \includegraphics[width=\textwidth]{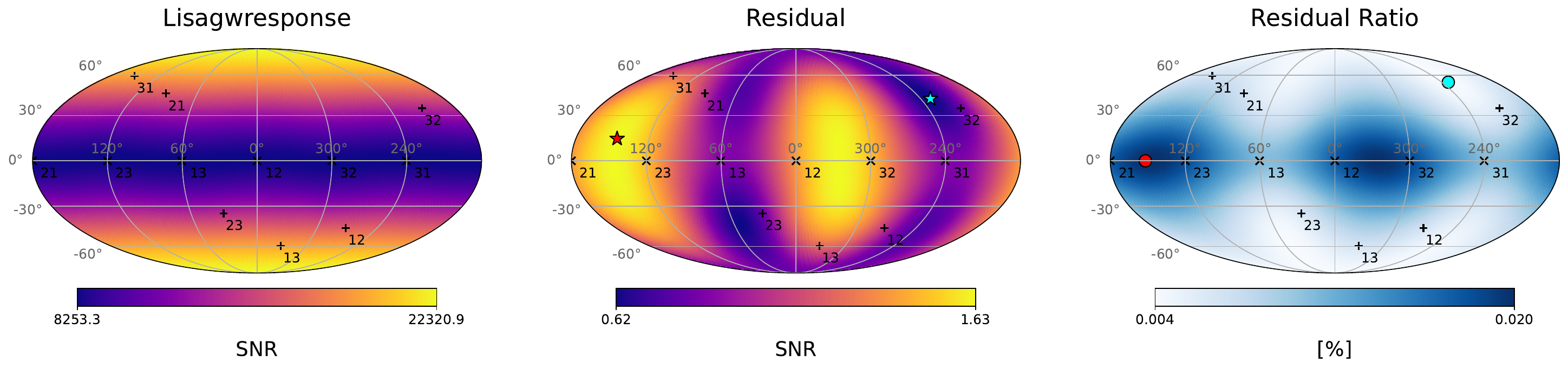}
    \caption{Sky-location dependence for a MBHB with redshifted total mass $M_\mathrm{tot} = \SI{5e6}{\mathrm{M}_\odot}$ at $z = 1$, with $q = 1$, $\vec\chi_1 = \vec\chi_2 = 0.7\hat z$, and $\iota = \pi/6$. We fix the polarization angle to $\psi = \pi/4$. The left sky-map shows the baseline SNR from \texttt{lisagwresponse}; the middle sky-map shows the residual SNR computed from the baseline response subtracted from the modified response; the right sky-map shows the ratio (residual SNR divided by baseline SNR). The reference plane is the LISA plane at time of merger. The labeled crosses ($\times$) indicate the directions of the link vectors $\hat{n}^{(0)}_{IJ}$ and the labeled plus signs ($+$) indicate the direction of the relative velocities $\Delta \vec{\beta}_{IJ}=\vec{\beta}_I-\vec{\beta}_J$. The red and cyan star indicate the maximum and minimum residual, respectively, and the red and cyan dot indicate the maximum and minimum ratio, respectively. The maps were generated using \texttt{Healpy} with $N_\mathrm{pix} = 768$ and smoothed with a symmetric Gaussian beam via the standard \texttt{smoothing} function~\cite{gorskiHEALPixFrameworkHighResolution2005, zoncaHealpyEqualArea2019}. Healpy partitions the sky into equal-area pixels.}
    \label{fig:Skymap_LISAplane}
\end{figure*}

\begin{figure*}
    \centering
    \includegraphics[width=\textwidth]{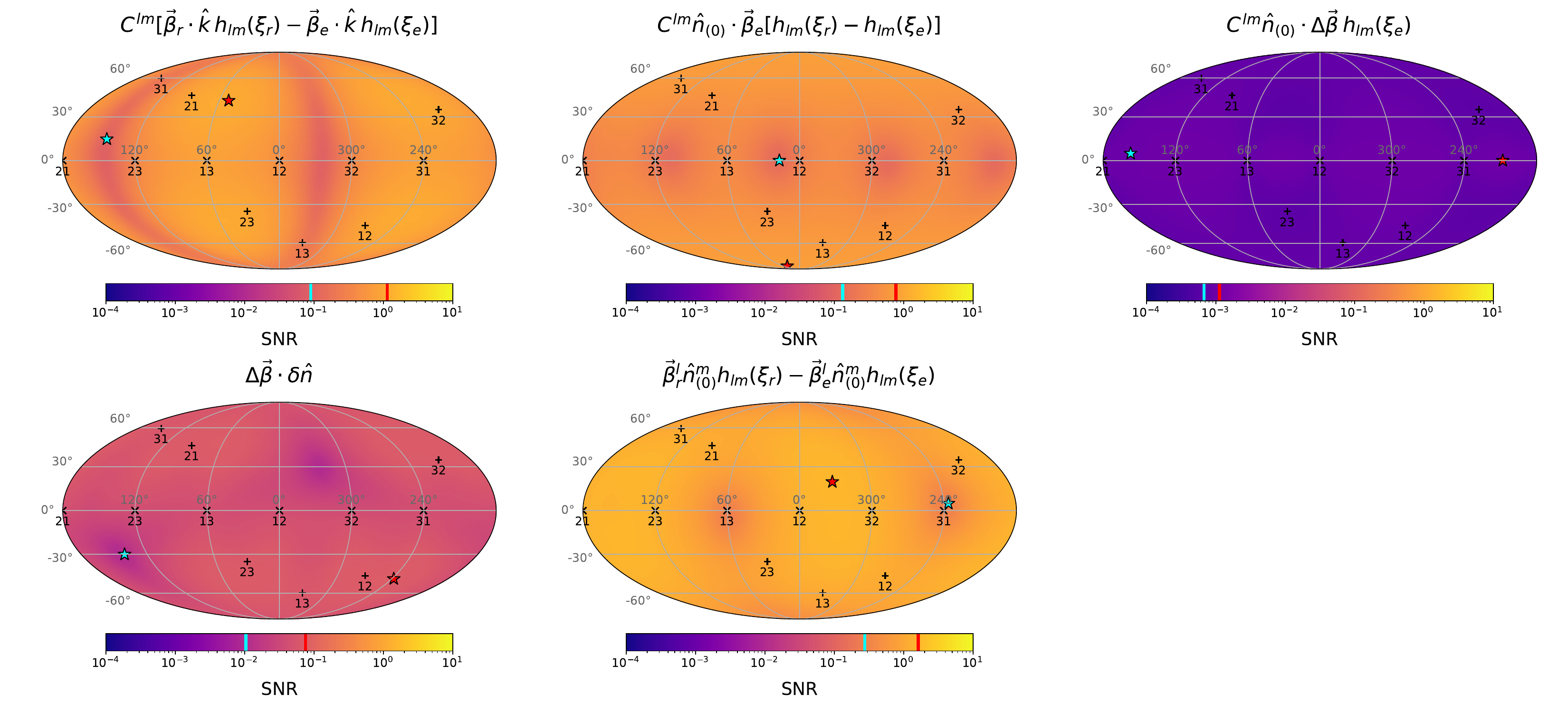}
    \caption{Sky-location dependence of the each individual corrections terms in Eq.~\eqref{eq:full_response}, using the same source parameters and polarization as in Fig.~\ref{fig:Skymap_LISAplane}. The reference plane is the LISA plane at time of merger. The labeled crosses ($\times$) indicate the directions of the link vectors $\hat{n}^{(0)}_{IJ}$ and the labeled plus signs ($+$) indicate the direction of the relative velocities $\Delta \vec{\beta}_{IJ}=\vec{\beta}_I-\vec{\beta}_J$. All panels use the same color scale. The red and cyan star indicate the maximum and minimum residual, respectively, for each term. This maximal and minimal SNR value is also indicated in the colorbar for each individual sky-map. In the titles, we use the shorthand $C^{lm} \equiv \hat{n}_{(0)}^l \hat{n}_{(0)}^m / (2[1 - \hat{k} \cdot \hat{n}_{(0)}])$ and $\Delta \beta \equiv\beta_r-\beta_e$. The sky-maps were generated with $N_\mathrm{pix} = 768$ and upsampled using \texttt{Healpy}'s \texttt{smoothing} function.}
    \label{fig:Skymap_Terms_LISAplane}
\end{figure*}

As seen in Eqs.~\eqref{eq:antenna_nn} and \eqref{eq:antenna_bn}, the detector response depends on the orientation of LISA relative to the GW propagation direction $\hat{k}$. We therefore begin by investigating the sky-location dependence of the total SNR. To enhance the visibility of velocity-dependent effects, we consider a high-SNR MBHB with redshifted total mass $M_\mathrm{tot} = \SI{5e6}{\mathrm{M}_\odot}$ at $z = 1$. The remaining source parameters, including inclination, polarization, and mass ratio, were chosen to  to construct a ``worst-case'' scenario, where the expected velocity-dependent corrections are maximized.

The results are shown in Fig.~\ref{fig:Skymap_LISAplane}. The left panel displays the baseline SNR computed using \texttt{lisagwresponse}. As expected from the antenna pattern of Eq.~\eqref{eq:antenna_nn}, LISA’s sensitivity varies significantly across the sky, exhibiting a band of reduced response near the plane spanned by the link vectors $\hat{n}_{(0)}$. The middle panel shows the residual SNR, obtained by subtracting the baseline frequency-domain TDI variables from those computed with the modified response. The residual exhibits a structured pattern of hot and cold spots across the sky. However, because the total correction arises from several competing terms in Eq.~\eqref{eq:response}, each with distinct geometric dependencies, it is not straightforward to attribute these features to specific link or velocity directions. The right panel shows the fractional residual, defined as the residual SNR divided by the baseline SNR. The largest relative deviations occur close to the LISA plane at merger, where the baseline response is suppressed. Nevertheless, the overall effect remains small: the maximal residual SNR reaches only $\sim 2$, while fractional differences remain below $0.020\%$. The corrections reach moderate values that may influence parameter inference.

To disentangle the origin of these features, Fig.~\ref{fig:Skymap_Terms_LISAplane} shows sky-maps of the residual SNR associated with each individual velocity-dependent correction term in Eq.~\eqref{eq:full_response}, evaluated for the same source configuration. Each contribution exhibits a distinct angular dependence and characteristic amplitude. For this system, the localized redshift, the point-ahead correction proportional to the leading response times $\hat{n}_{(0)}\cdot\vec{\beta}_e$, and the spacecraft-motion Doppler shift provide the dominant contributions, yielding residual SNRs of order unity. The propagation Doppler shift, $\Delta\vec{\beta}\cdot\delta\hat{n}$, contributes at a moderate level, while the point-ahead correction proportional to $\hat{n}{(0)}\cdot\Delta\vec{\beta}$ is negligible, consistent with the small relative spacecraft velocities, $\Delta\beta \sim 10^{-6}$.

The geometric origin of several features can now be understood more clearly. The reduced-response band in Fig.~\ref{fig:Skymap_LISAplane} occurs when $\hat{k}$ lies approximately in the plane spanned by the link vectors $\hat{n}_{(0)}$, in agreement with the antenna structure of Eq.~\eqref{eq:antenna_nn}. The spacecraft velocities are also nearly confined to this plane, whereas the relative velocities $\Delta\vec{\beta}$ predominantly point out of the plane, explaining some of the contrasting angular behavior between correction terms. For example, both Doppler corrections (the two bottom panels in Fig.~\ref{fig:Skymap_Terms_LISAplane}) display a symmetry with respect to the plane spanned by the relative velocities. One correction is enhanced when $\hat{k}$ lies close to this plane, whereas the other is largest when $\hat{k}$ points approximately perpendicular to it.

For completeness, sky-maps expressed in the original Equatorial coordinates are provided in App.~\ref{app:Extra}; Figs.~\ref{fig:Skymap_Equatorial} and \ref{fig:Skymap_Terms_Equatorial} correspond to rotated versions of Figs.~\ref{fig:Skymap_LISAplane} and \ref{fig:Skymap_Terms_LISAplane}, respectively, shown in equatorial longitude $\alpha$ and equatorial latitude $\delta$.

\begin{figure*}[t]
    \centering
    \includegraphics[width=\textwidth]{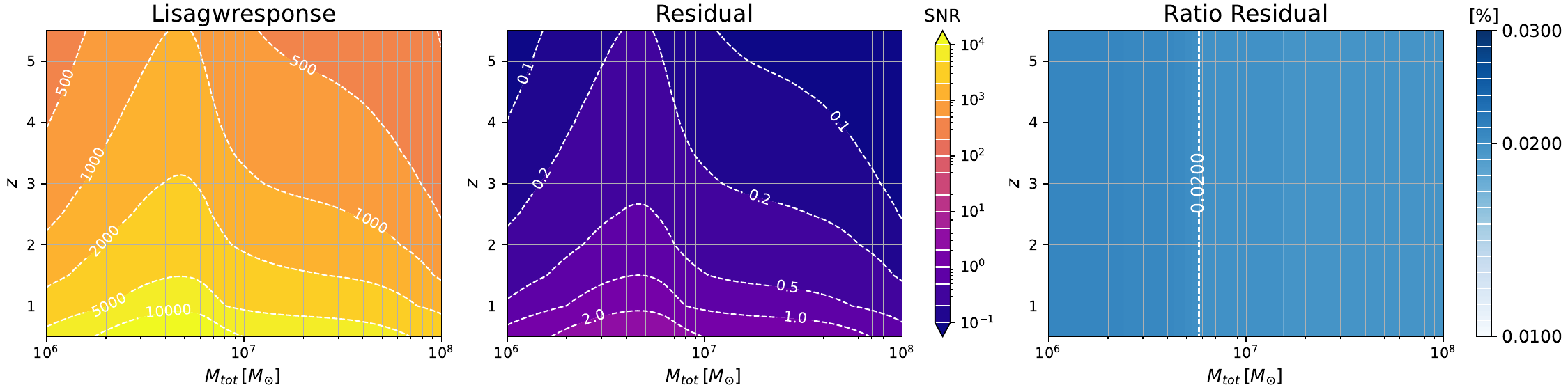}
    \caption{Total (redshifted) mass and redshift dependence of the response modifications. The left panel shows the baseline SNR from \texttt{lisagwresponse}; the middle panel the residual obtained by subtracting the baseline from the modified response; the right panel shows the ratio (residual SNR divided by baseline SNR). Simulations have been generated using parameters $q = 1$, $\vec\chi_1 = \vec\chi_2 = 0.7\hat z$, and $\iota = \pi/6$. The sky-position has been fixed at the maximum ratio depicted in Fig.~\ref{fig:Skymap_LISAplane} with polarization angle $\psi = \pi/4$.}
    \label{fig:LogMvsZ}
\end{figure*}

The modified response contains several competing contributions that exhibit different frequency dependencies. To probe the broader parameter space, we examine how the SNR residual depends on the total (redshifted) mass $M_{\mathrm{tot}}$ and the redshift $z$. The sky position is fixed at the location of the maximum ratio, indicated by the red dot in the right panel of Fig.~\ref{fig:Skymap_LISAplane}. While the precise location of the maximum ratio drifts slightly with mass, it remains close to the position shown in Fig.~\ref{fig:Skymap_LISAplane} (see Fig.~\ref{fig:Skymaps_Ms_Ratio}). All other parameters, such as the mass ratio and inclination, are kept fixed relative to the previous analysis.

The results of varying the mass and redshift are shown in Fig.~\ref{fig:LogMvsZ}. Compared to similar studies in the literature (e.g.~\cite{pitteDetectabilityHigherHarmonics2023, inchauspeMeasuringGravitationalWave2025}), our choice of source parameters probes a relatively optimistic region of parameter space, with favorable high-SNR observations, which provide a sensitive test of the impact of velocity-dependent corrections. The residual SNR is shown in the middle plot and follows the expected trend that sources with larger baseline SNRs also exhibit larger residuals. Residual SNRs approach order unity primarily for MBHBs at low redshift, $z \leq 1$, corresponding to the most favorable observational scenarios, although such systems are expected to be comparatively rare \cite{salcidoMusicHeavensGravitational2016, katzEvaluatingBlackHole2019, katzProbingMassiveBlack2020}. In the right panel, the response ratios remain remarkably stable across both redshift and total mass, consistently staying around $0.02\%$, suggesting that the relative impact of the correction is largely insensitive to source parameters.

\section{\label{sec:discussion} Conclusion and Discussion}

We have derived and implemented a velocity-dependent correction to the LISA response function in Eq.~\eqref{eq:full_response} arising from spacecraft movement during the light travel time. Using a perturbative expansion at linear order in both the GW amplitude $h$ and the spaceracft velocity $\beta$, we identified multiple subleading terms in the response function and assessed their effect on the detected SNR for MBHBs. This includes localized redshifts at both the emitting and receiving spacecraft, point-ahead corrections due to the emission time being unknown, and a correction to the laser's Doppler shift due to the line-of-sight being modulated and shifts in the spacecraft velocity due to the GW. Each of these contributions, has a distinctive sky-localization and frequency dependence.  

Our results confirm earlier suspicions in the literature~\cite{rubboForwardModelingSpaceborne2004} that these velocity corrections are small, even for the loudest MBHBs in the LISA band. We find residual SNRs of at most $\sim 2$ for low-redshift, equal-mass binaries with intrinsic SNRs of order $10^4$, consistent with corrections scaling as $\beta \sim 10^{-4}$. Fractionally, the largest differences reach $\SI{0.02}{\%}$, and this ratio remains remarkably stable across the full range of masses considered. While larger residuals may occur in specific sky locations or for certain parameters, we do not expect them to exceed a residual of $\sim 2$, since the slices of the parameter space considered in this study cover optimistic choices to maximize the effect of the velocity corrections.

With residual SNRs at most of order unity, the velocity-induced modulations are comparable in magnitude to the noise, but they are not necessarily negligible. The required precision for MBHB analysis in LISA is extremely high, and corrections at the $10^{-4}$ level are comparable to the accuracy of state-of-the-art numerical relativity waveforms~\cite{scheelSXSCollaborationsThird2025, purrerGravitationalWaveformAccuracy2020}. Based on SNR alone, we cannot conclude whether velocity corrections will influence parameter estimation. A rigorous assessment will require a full Bayesian analysis. This comes with caveats: generating the modified time-domain response is computationally demanding taking seconds to minutes per event (a factor $\sim 2$ slower than \texttt{lisagwresponse}), which makes large-scale Monte Carlo studies challenging. Implementing this in the frequency domain is challenging, since the fourth correction term depends solely on $h_{ij}(\xi_e)$ rather than differences in $h$, preventing a straightforward Fourier treatment (see also App.~\ref{app:LF}). However, the bias can still be assessed in future studies by generating data using the full response and then analyzing it using the usual transfer function used in current parameter-estimation tools~\cite{marsatExploringBayesianParameter2021} to circumvent computational issues.

Beyond their modest contribution to the overall SNR, velocity-dependent corrections encode additional information about the detector geometry. The distinct angular structure of the individual terms may help improve sky localization and resolve sky-position degeneracies~\cite{marsatExploringBayesianParameter2021}. In particular, the Doppler modulation terms breaks the reflection symmetry with respect to the LISA plane. Although the overall magnitude of the corrections remains small, we find that their relative impact is remarkably stable across the MBHB parameter space explored here, with fractional differences remaining at the level of $\sim 0.02\%$ at favorable sky-positions. This robustness suggests that velocity-dependent effects may provide a systematic source of geometric information even when their contribution to the total SNR is subdominant.

High-mass MBHBs are attractive systems for future investigation. Because these binaries spend comparatively little time in the LISA band, they benefit less from the orbital modulation that ordinarily aids sky-localization, potentially increasing the relative importance of additional directional information encoded in the detector response. At the same time, their short signal duration reduces the computational cost associated with including the full velocity-dependent response in parameter estimation. Incorporating these corrections into inference studies therefore remains a promising avenue for future work.

For simulations and mock data production, we recommend incorporating the full response. In this context the goal is to generate datasets faithful to future measurements, and the velocity corrections bring the LISA response to the same accuracy level as current NR waveforms. Since the response only needs to be generated once per dataset, computational speed is not a limiting factor. 

Finally, velocity corrections will become more relevant for other space-based GW detectors and future space-based missions beyond LISA. For example, the proposed LISAmax concept envisions three spacecraft distributed in an equilateral triangle centered around the sun with arms of 259 million km, more than two orders of magnitude longer than LISA’s, operating in the \SI{}{\micro\hertz} band and offering two orders of magnitude greater sensitivity below \SI{e-3}{\hertz}~\cite{martensLISAmaxImprovingLowfrequency2023}. In such a mission, spacecraft would still orbit at velocities $\beta \sim 10^{-4}$ in the SSB frame, and relative velocities $\Delta\vec{\beta}={\beta}_r-{\beta}_e$ would persist at the same order $\Delta\beta\sim 10^{-4}$ due to the constellation being confined to the ecliptic plane. In this case, all four velocity-dependent correction terms would likely exceed the noise and therefore be essential for accurate modeling. Moreover, because LISAmax would lack seasonal Doppler modulation, sky-localization would be particularly challenging without them. For such next-generation detectors, incorporating the full response including velocity corrections is not optional but a requirement for success.

\section*{Acknowledgments}

We would like to thank Jann Zosso for raising some insightful points. Tom van der Steen acknowledges funding from the Research Foundation - Flanders (FWO, Fonds Wetenschappelijk Onderzoek) through a PhD Fellowship under FWO Grant No. 1125026N. Tom van der Steen, Henri Inchausp\'e and Thomas Hertog acknowledge support from the Flemish inter-university project IBOF/21/084 and thank the Belgian Federal Science Policy Office (BELSPO) for the provision of financial support in the framework of the PRODEX Programme of the European Space Agency (ESA) under contract number PEA4000144253. The resources and services used in this work were provided by the VSC (Flemish Supercomputer Center), funded by the Research Foundation - Flanders (FWO) and the Flemish Government.

\bibliographystyle{apsrev4-2}
\bibliography{Bibliography}

\appendix

\section{Two time scales approximation to the equations of motion}\label{app:motion}
In this appendix, we will discuss the solution to Eq.~\eqref{eq:acc_pert} and we will focus mainly on the velocity perturbations, i.e. at $V_x=X^{\prime\, 1}_{(1)}$, $V_y=X^{\prime\,2}_{(1)}$ and $V_z=X^{\prime\,3}_{(1)}$. For the sake of this theoretical study, we will consider a monochromatic GW as a source (e.g. an inspiraling binary system far from merger, such as white dwarf binaries in our galaxy). More general GW sources can be considered by convoluting our results. The GW source is therefore parametrized by seven parameters: the frequency $\fg$, the initial phase $\phi_0$, the sky-location provided by the right ascension $\alpha$ and the declination $\delta$, the polarization angle $\psi$, the amplitude $A$ and the inclination $\iota$.

Let us introduce the following 6-dimensional vector to describe the solution to Eq.~\eqref{eq:acc_pert}:
\begin{equation}
    \bm y =\left (\oo X^1_{(1)}, \oo X^2_{(1)}, \oo X^3_{(1)},X^{\prime\,1}_{(1)}, X^{\prime\, 2}_{(1)}, X^{\prime\, 3}_{(1)}\right),
\end{equation}
where $\oo=2\pi \fo$. In the case of LISA, we have $\oo=2\pi/\SI{1}{yr}$. The prefactor $\oo$ is introduced so that all components of $\bm y$ carry units of velocity. Using $\bm y$, Eq.~\eqref{eq:acc_pert} can then be written as 
\begin{equation}\label{eq:yprime}
     y^{\prime\,i} = F^{i}_{\ j}y^j + \delta a^i\, , 
\end{equation}
where $F^{i}_{\ j}$ is a $6\times 6$ matrix defined by
\begin{equation}\label{eq:F}
    F^{i}_{\ j} = \begin{pmatrix} \mathbf 0_{3\times3} & \oo \mathbf I_{3\times3}  \\ \tilde {\bm F} & \mathbf 0_{3\times3}   \end{pmatrix}\, ,
\end{equation}
where $\mathbf 0_{3\times3}$ is a $3\times3$ vanishing matrix, $\mathbf I_{3\times3}$ is the $3$-dimensional identity matrix and $\tilde {\bm F}$ is a 3$\times$3 matrix defined by
\begin{align}
    \tilde F^{i}_{\ j} &=\frac{1}{\oo} \left.\frac{\partial a^i}{\partial X^j}\right|_{X^i_{(0)}} \nonumber\\
    &\approx -\frac{1}{\oo}\frac{GM_\odot}{\left|\vec X_{(0)}\right|^3}\left(\delta^i_j - 3 \frac{X^i_{(0)}X_{j(0)}}{\left|\vec X_{(0)}\right|^2}\right)\, ,
\end{align}
where we have included only the leading contribution, the Newtonian acceleration due to the Sun. The perturbed acceleration $\delta a^i$ which appears in Eq.~(\ref{eq:yprime}) is a $6$-dimensional vector defined as
\begin{equation}
    \delta a^i = \begin{pmatrix} 
    \mathbf{0}_3 \\ 
    -h^\prime_{1j}X^{\prime\,j}_{(0)} \\
    -h^\prime_{2j}X^{\prime\,j}_{(0)}\\ 
    -h^\prime_{3j}X^{\prime\,j}_{(0)}  
    \end{pmatrix}\, .
\end{equation}

\begin{figure*}
    \centering
    \includegraphics[width=0.8\textwidth]{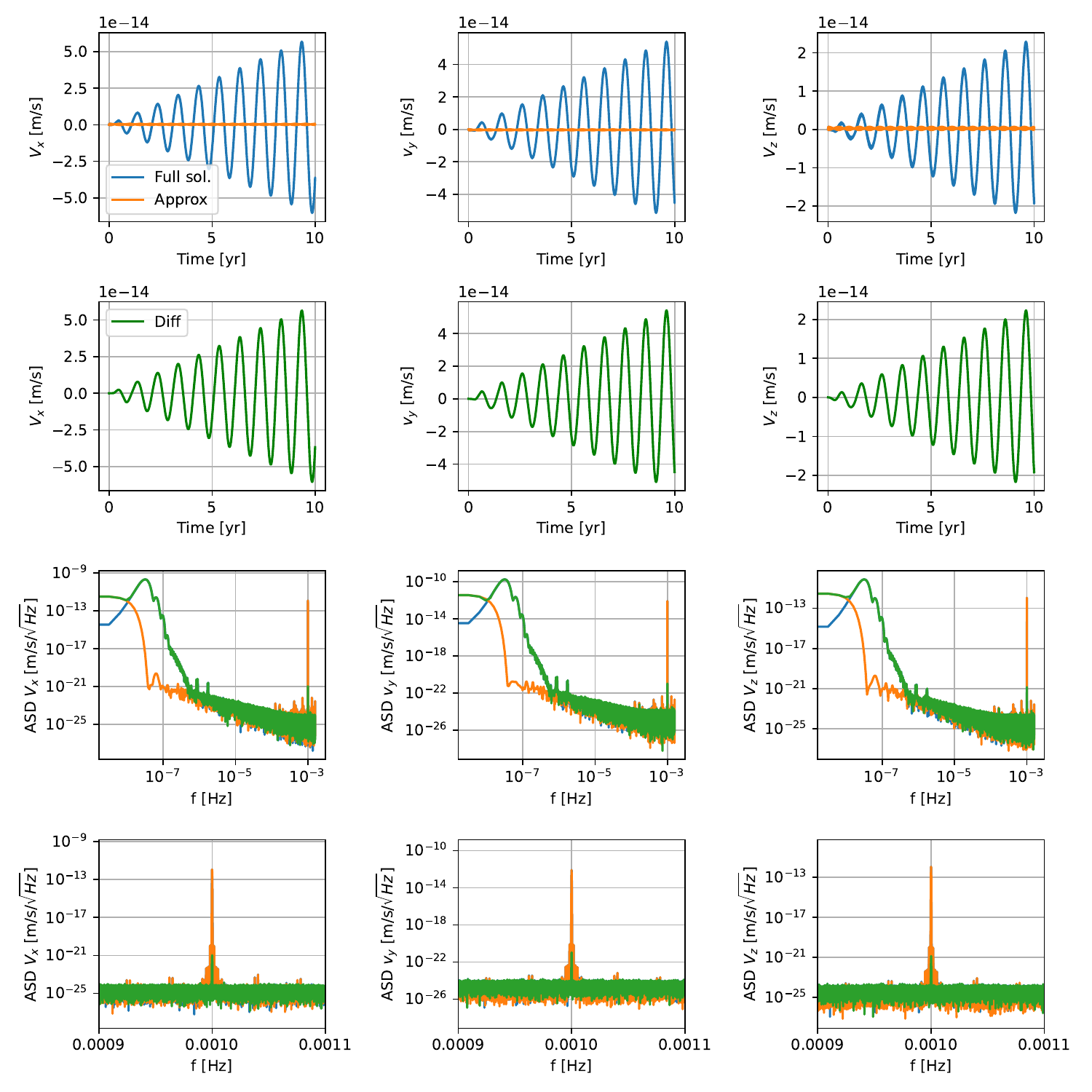}
    \caption{Top panels: in blue is presented the velocity of LISA spacraft 1 for the full solution to Eq.~(\ref{eq:yprime}) while the cyan curves correspond to the solution of the same equation when the $F^i_{\ j}$ matrix is neglected. Middle panels: the difference between the full solution and the approximate solutions. The two bottom panels present the ASD of the same curves (the last line being a zoom version around the GW frequency). The columns correspond to the 3 coordinates. The GW used in this simulation has a strain amplitude of $10^{-20}$ and a frequency of $10^{-3}$ Hz. As a reminder, the peak of the ASD of a harmonic signal of amplitude $A$ is given by $A\sqrt{T_\mathrm{obs}/2}$ where $T_\mathrm{obs}$ is the observation time.}
    \label{fig:sol_motion}
\end{figure*}

Fig.~\ref{fig:sol_motion} presents the result of the integration of Eq.~\eqref{eq:yprime} for SC1 over a 10 year period for a monochromatic GW of frequency $10^{-3}$ Hz using zeroth-order spacecraft orbits provided by the \texttt{LISAOrbits} Python package~\cite{bayleLISAOrbits2022, martensTrajectoryDesignESA2021}, using the numerically optimized ESA trailing orbit configuration. On the top panels from Fig.~\ref{fig:sol_motion}, the blue curves correspond to the solution of Eq.~\eqref{eq:yprime}, while the cyan curve corresponds to the solution of the same equation where the $F^i_j$ contribution is neglected, i.e. the cyan curve corresponds to a solution of Eq.~\eqref{eq:ddotxi}. The difference between these two curves are presented in the middle panels (green curves). The ASD of all these curves are presented in the bottom panels. At the GW frequency $\fg=\SI{e-3}{Hz}$, we can see that the $F^i_j$ contribution (green) is suppressed by about nine orders of magnitude compared to the $\delta a^i$ contribution (cyan). This corresponds to a factor $\fg^2/\fo^2\sim10^{-9}$ as we will see below. The important result conveyed by this numerical analysis is that, in case the orbital frequency is way smaller than the GW frequency, the full solution to Eq.~(\ref{eq:yprime}) can be expressed, to sufficiently good approximation, as the sum of two terms: (i) one term that oscillates at high frequencies (the cyan curves which correspond to the approximation used in this paper) and (ii) an out-of-band component (i.e. a component which oscillates at small frequencies $\sim \fo$, as depicted by the green curves on Fig.~\ref{fig:sol_motion}). This shows that one can safely use Eq.~(\ref{eq:ddotxi}) for modeling the in-band contribution from the GW.

Let us now strengthen this conclusion by providing an analytical argument. The solution to Eq.~(\ref{eq:yprime}) can be obtained by using the fundamental matrix, see e.g. \cite{2004sod..book.....T,foster:2025}. The fundamental matrix $\Phi^i_{\ j}$ is a 6$\times$6 matrix defined by
\begin{equation}
    \frac{\mathrm{d}}{\mathrm{d}t} \Phi^i_{\ j}(t) = F^i_{\ k}(t) \Phi^k_{\ j }(t), \qquad \Phi^i_{\ j}(t = 0) = \delta^i_j\, .
\end{equation}
For completeness, let us mention that the inverse of the fundamental matrix follows a similar equation
\begin{align}\label{eq:phi_inv}
\frac{\mathrm{d}}{\mathrm{d}t} \left[\Phi^{-1}\right]^i_{\ j}(t)& = -\left[\Phi^{-1}\right]^i_{\ k}(t)F^k_{\ j}(t) \, ,\\
\left[\Phi^{-1}\right]^i_{\ j}(t = 0) &= \delta^i_j\, . \nonumber
\end{align}
Once the fundamental matrix has been evaluated, the solution to Eq.~(\ref{eq:yprime}) is then given by
\begin{equation}\label{eq:sol_from_Phi}
    y^i(t) = \Phi^{i}_{\ j}(t) \int_0^{t} \mathrm{d}\tau \, \left[\Phi^{-1}\right]^j_{\ k}(\tau) \delta a^k(\tau)\, .
\end{equation}

\begin{figure*}
    \centering
    \includegraphics[width=\textwidth]{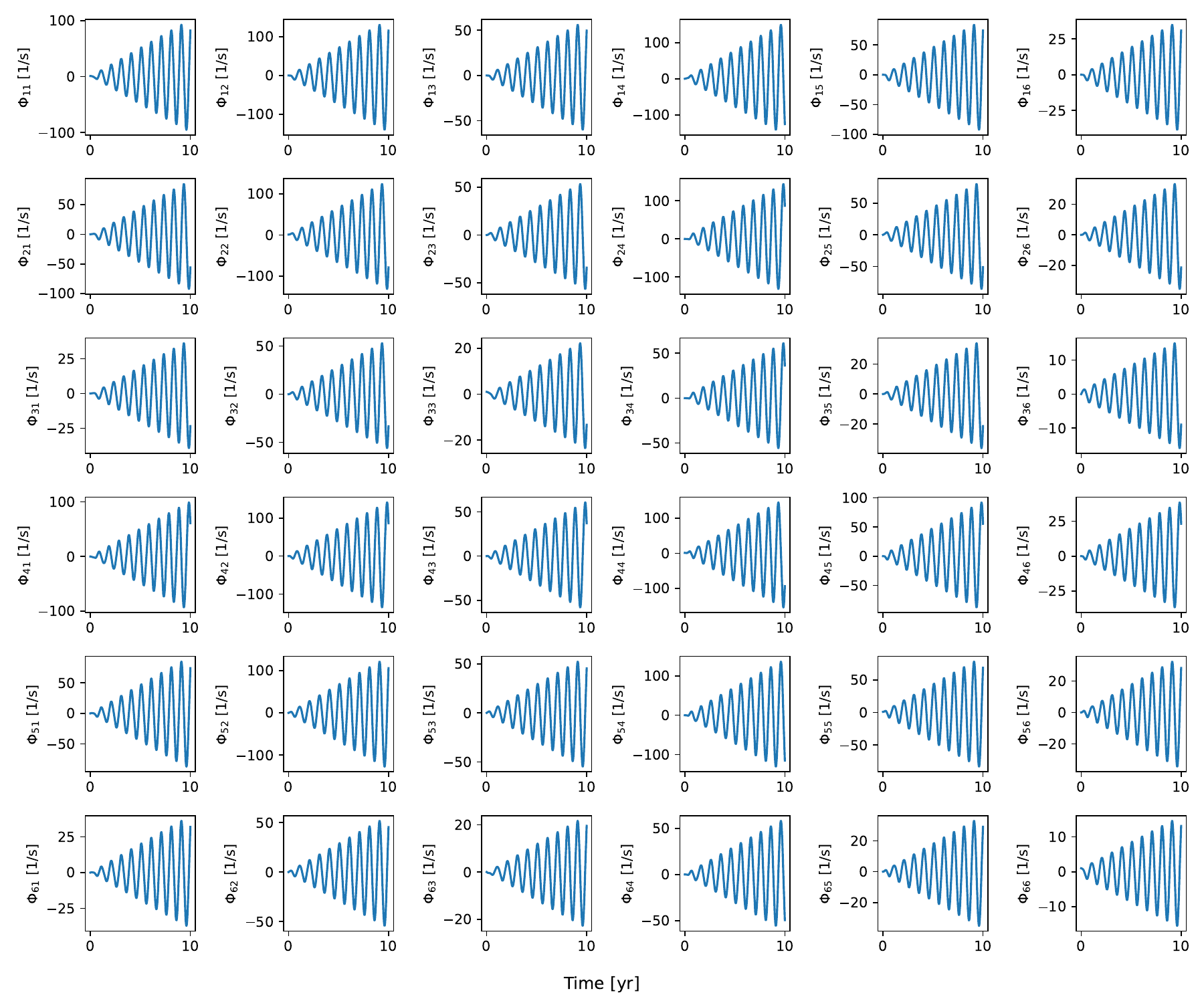}
    \caption{fundamental matrix $\Phi_{\mu\nu}$ for spacecraft 1 for a period of 10 years using the ESA optimized trailing orbits. All the components of $\Phi_{\mu\nu}$ have an oscillatory behavior at the orbital period with an amplitude that grows with time.}
    \label{fig:Phi}
\end{figure*}

The $\tilde {\bm F}$ matrix contains terms that oscillate mainly at twice the orbital frequency (with other harmonics as well) with an amplitude $\sim \oo$. The fundamental matrix, which depends only on the zeroth order trajectory is shown on Fig.~\ref{fig:Phi} for illustrative purpose. This matrix behaves approximately as $\bm \left(I_{6\times6} + \frac{3}{2}\oo t \bm \Lambda\right) \bm P(t)$ where $\bm \Lambda$ is a 6$\times6$ constant matrix and $\bm P(t)$ is a 6$\times6$ periodic matrix (whose period is the orbital period), in line with Floquet's theorem \cite{barone1977floquet}. The inverse of the fundamental matrix presents similar characteristics. 

In order to validate our approximation, let us integrate Eq.~(\ref{eq:sol_from_Phi}) by parts using Eq.~(\ref{eq:phi_inv}) to get
\begin{align}\label{eq:yi_sol}
    y^i(t) = &\int_0^t \mathrm{d}\tau\ \delta a^i(\tau) \\
    &\quad + \Phi^i_j(t) \int_0^t \mathrm{d}\tau \left[\Phi^{-1}\right]^i_{\ l}(\tau)F^l_{\  k}(\tau)\int_0^\tau \mathrm{d}s\ \delta a^k(s) \, .  \nonumber
\end{align}
The first term on the right hand side corresponds to the approximate solution, i.e. to the cyan curves from Fig.~\ref{fig:sol_motion} while the second term corresponds to the neglected contribution, i.e. the green curves from Fig.~\ref{fig:sol_motion}.

The $\delta a^i$ term oscillates at a frequency $\sim \fg$ and has an amplitude of the order of $\sim 2\pi \fg v h$ where $v$ is the typical velocity of the spacecraft ($\sim 30$ km/s) and $h$ is the GW strain amplitude ($h=10^{-20}$ in the numerical results presented in this section)\footnote{In reality, $\delta a^i$ has a frequency content which is centered on $\fg$ and which contains several sidebands $\fg \pm n \fo$ with $n$ an integer due to the orbital motion. This complexity does not change the reasoning.}. The term $\int \mathrm d\tau  a^i$ therefore contains terms that oscillate at $\sim \fg$ and a DC term, both of them being of amplitude $\sim vh$ ($\sim \SI{3e-16}{m/s}$ using the numerical values used in this section). 

Let us now focus only on the in-band (fast oscillating term). Using Eq.~\eqref{eq:F}, one can see that
\begin{align}
    F^l_{\ k}(\tau) \int_0^\tau \mathrm{d}s \: \delta a^k(s) 
    &= \oo \begin{pmatrix}
        \int \delta a^1 \\ 
        \int \delta a^2 \\ 
        \int \delta a^3 \\
        \, \mathbf{0}_3
   \end{pmatrix} \nonumber\\
   &\approx \oo v h \begin{pmatrix}
        \osc(\tau) \\ \bm 0_3
    \end{pmatrix}\, , 
\end{align}
where $\osc(\tau)$ denotes a vector of three terms that all oscillate at frequencies close to $\fg$ and with amplitude $\mathcal O(1)$. One can therefore write the second term from Eq.~(\ref{eq:yi_sol}) and integrate by part using Eq.~(\ref{eq:phi_inv})
\begin{widetext}
\begin{subequations}
\begin{align}
    \oo v h \ \bm \Phi(t)\int_0^t \mathrm{d}\tau  \ \bm \Phi^{-1}(\tau) &\begin{pmatrix}
        \osc(\tau) \\ \bm 0_3 
    \end{pmatrix}
    = \frac{\oo}{\og} v h \ \begin{pmatrix}
        \osc(t) \\ \bm 0_3 
    \end{pmatrix} - \frac{\oo}{\og} v h \ \bm \Phi(t) \int_0^t \mathrm{d}\tau  \  \bm \Phi^{-1}(\tau) \bm F(\tau) \begin{pmatrix}
        \osc(\tau) \\ \bm 0_3  
    \end{pmatrix}\,  \nonumber\\
    =&\frac{\oo}{\og} v h \ \begin{pmatrix}
        \osc(t) \\ \bm 0_3 
    \end{pmatrix}- \frac{\omega_{\mathrm{orb}}^2}{\og} v h \ \bm \Phi(t) \int_0^t \mathrm{d}\tau  \  \bm \Phi^{-1}(\tau) \begin{pmatrix}
         \bm 0_3   \\ \osc(\tau) 
    \end{pmatrix}\nonumber \\
    =&\frac{\oo}{\og} v h \ \begin{pmatrix}
        \osc(t) \\ \bm 0_3 
    \end{pmatrix} - \frac{\omega_\mathrm{orb}^2}{\og} v h \int_0^t\mathrm d\tau \begin{pmatrix}
         \bm 0_3   \\ \osc(\tau) 
    \end{pmatrix} \nonumber\\
    &\qquad \qquad + \frac{\omega_\mathrm{orb}^2}{\omega_\mathrm{GW}^2} v h \bm \Phi(t)  \int_0^t\mathrm d\tau \bm \Phi^{-1}(\tau)\bm F(\tau) \begin{pmatrix}
         \bm 0_3   \\ \osc(\tau) 
    \end{pmatrix} \nonumber\\ 
    =& \frac{\oo}{\og} v h \ \begin{pmatrix}
        \osc(t) \\ \bm 0_3 
    \end{pmatrix} - \frac{\omega_\mathrm{orb}^2}{\omega_\mathrm{GW}^2} v h\begin{pmatrix}
         \bm 0_3   \\ \osc(\tau) 
    \end{pmatrix}+\mathcal O\left(\frac{\omega_\mathrm{orb}^3}{\omega_\mathrm{GW}^3}\right) \, .\label{eq:ap_neglect}
\end{align}
\end{subequations}
\end{widetext}
We integrated by part for the first equality. To go from the first to second line, we used Eq.~(\ref{eq:F}). To go from the second line to the third, we integrated by parts. In conclusion, we have shown that the leading in-band contribution for the velocity (i.e. the last three components) will have an amplitude of $\frac{\omega_\mathrm{orb}^2}{\omega_\mathrm{GW}^2} v h$ ($\sim 10^{-9} vh \sim\SI{3e-25}{m/s^2}$ in the simulation presented in this section) which is fully negligible compared to the first term of Eq.~(\ref{eq:yi_sol}) which has an amplitude of $\sim vh$. This demonstrates that when we are interested only in the in-band component and when the orbital frequency is much larger than the orbital frequency, one can safely neglect the second term from Eq.~(\ref{eq:yi_sol}), which corresponds to neglecting the $F^i_{\ j}y^j$ term in Eq.~(\ref{eq:yprime}) or the $\left.\frac{\partial a^i}{\partial X^j}\right|_{X^i_{(0)}} X^j_{(1)}$ term in Eq.~(\ref{eq:acc_pert}). This analytical reasoning validates the numerical example presented in the beginning of this appendix and summarized in Fig.~\ref{fig:sol_motion}.

For the sake of completeness, let us note that the properties of the fundamental matrix show that the out-of-band contribution to Eq.~(\ref{eq:yi_sol}) oscillates at the orbital frequencies with an amplitude which grows as $\sim 3\oo t vh$ (which for the example presented in Fig.~\ref{fig:sol_motion} corresponds to $\sim \SI{6e-14}{m/s^2}$, in good agreement with the green curves).

\subsection{The link vector}\label{app:pointing}
In the previous subsection, we focused on the impact of the GW on the test mass velocity, which is given by Eq.~\eqref{eq:GW_velocity}. Let us now show how this propagates into the link vector. First of all, let us integrate Eq.~\eqref{eq:GW_velocity} by parts:
\begin{equation}\label{eq:position}
    X^i_{(1)}(t) = - \frac{1}{c}\tilde H_{ij}(t) X^{\prime \, j}_{(0)}(t)  + \frac{1}{c}\int_0^t \mathrm{d}\tau \tilde H_{ij}(\tau) X^{\dprime \, j}_{(0)}(\tau)\, , 
\end{equation}
where $\tilde H_{ij}$ is the integral of the metric over the zeroth-order trajectory, i.e. 
\begin{equation}
    \tilde H_{ij}(t)=\int_0^{ct} \mathrm{d}(c\tau) \:  h_{ij}\left(\tau,X^j_{(0)}(\tau)\right) \, .
\end{equation}
In Eq.~\eqref{eq:position}, the last term is of order $\mathcal O(f_\mathrm{orb}/f_\mathrm{GW})$ smaller than the first one and can safely be neglected. The first term is of order $\mathcal O(hv/\omega_\mathrm{GW})$ in agreement with the first term of Eq.~\eqref{eq:ap_neglect}.

The link vector is defined by Eq.~\eqref{eq:pointing_vec}. A perturbative expansion of this unit vector leads to
\begin{equation}
    \hat{n}_{(0)} = \frac{\vec{X}_{r}^{(0)}(t_r) -\vec{X}_{e}^{(0)}(t_e^{(0)})}{L_{(0)}} +\hat{n}_{(1)} +\mathcal O (h^2) \, ,
\end{equation}
where $L_{(0)}=\left|\vec{X}_{r}^{(0)}(t_r) -\vec{X}_{e}^{(0)}(t_e^{(0)})\right|$ and
\begin{widetext}
\begin{equation}
    \hat{n}_{(1)} = \frac{\vec{X}_{r}^{(1)}(t_r) -\vec{X}_{e}^{(1)}(t_e^{(0)})}{L_{(0)}} - \frac{\left(\vec{X}_{r}^{(1)}(t_r) -\vec{X}_{e}^{(1)}(t_e^{(0)})\right)\cdot\left( \vec{X}_{r}^{(0)}(t_r) -\vec{X}_{e}^{(0)}(t_e^{(0)})\right)}{L_{(0)}^3}\left(\vec{X}_{r}^{(0)}(t_r) -\vec{X}_{e}^{(0)}(t_e^{(0)})\right)\, .
\end{equation}
\end{widetext}
Since $X^i_{(1)}\sim \mathcal{O}(hv/\omega_\mathrm{GW})$, one gets that 
\begin{equation}
    \hat{n}_{(1)}\sim \mathcal{O}\left(\frac{h}{\omega_\mathrm{GW}}\frac{\Delta v}{\Delta D}\right)\, ,
\end{equation}
where $\Delta v$ is the typical velocity difference between two test masses and $\Delta D$ their typical distance. In the case of LISA, the velocity difference of two test masses is typically of the order of $\Delta v\sim 500$ m/s, while the typical armlength $\Delta D\sim \SI{0.017}{AU}$, such that $\Delta v/\Delta D\sim \omega_\mathrm{orb}$. Hence, 
\begin{equation}
    \hat n_{(1)} \sim \mathcal O \left(h \frac{\omega_\mathrm{orb}}{\omega_\mathrm{GW}}\right)\, .
\end{equation}
This shows that the term $(\beta^i_{r,(0)}-\beta^i_{e,(0)}) \hat n^i_{(1)}$ which should also appear in Eq.~\eqref{eq:sc_acc_doppler} is of the order $\mathcal O(hv f_\mathrm{orb}/f_\mathrm{GW})$, i.e. a factor $f_\mathrm{orb}/f_\mathrm{GW}$ smaller than the leading term from Eq.~\eqref{eq:sc_acc_doppler}, which can be safely be neglected.

\section{\label{app:TTF} Time Transfer Function Derivation}

As an alternative to the Doppler tracking derivation for the light propagation derivation presented in Sec.~\ref{sec:light}, the modified response can also be obtained using the time transfer function (TTF) formalism \cite{poncin-lafitteWorldFunctionTime2004,*teyssandierGeneralPostMinkowskianExpansion2008,*hees:2014fk}. A perturbative post-Minkowskian (pM) development of the TTF is particularly useful since it expresses the light travel time using the unperturbed straight-line path between the emission and reception events instead of solving the full perturbed geodesic \cite{poncin-lafitteWorldFunctionTime2004,*teyssandierGeneralPostMinkowskianExpansion2008,*hees:2014fk}.

\subsection{Proper time versus coordination time}
In General Relativity, an observable (i.e. a quantity that can be measured) is expressed in terms of a quantity that is invariant under diffeomorphisms. In the case of LISA, this corresponds to the proper light-travel time (i.e. the light-travel time as measured by clocks) or the ratio of proper frequency. These observables are therefore based on proper-time, which, in the TT-gauge is related to coordinate time through
\begin{equation}
    \frac{\mathrm d\tau}{\mathrm dt}= 1 - \frac{V^2}{2c^2} + \mathcal O(V^2h/c^2)\, ,
\end{equation}
which means that the contribution of the GW to the evolution of proper time with respect to coordinate time is negligible and we can focus on the impact of the GW on the light-travel time expressed in terms of coordinate time only.

\subsection{Light propagation effects}
In this section, we will compute the impact of the gravitational wave on the coordinate light-travel time as a function (of the assumed known) receiver and emitter's trajectories.

\subsubsection{Static spacecraft}
First, let us revisit the calculation of the light travel time in the case where both spacecraft are at rest.

The TTF is defined as the coordinate light travel time, $t_r - t_e$, of a photon propagating from the emission point $\vec{X}_e$ to the reception point $\vec{X}_r$. It can be viewed either as a function of $(t_e, \vec{X}_e, \vec{X}_r)$ or of $(t_r, \vec{X}_e, \vec{X}_r)$, depending on wether the emission time $t_e$ or reception time $t_r$ is chosen to parameterize the system's evolution. Consistent with the earlier derivation, we assume the reception time is given and introduce the reception time transfer function:
\begin{equation}
    t_r-t_e=\mathcal{T}_r( \vec{X}_e,t_r,\vec{X}_r)\, .
\end{equation}

In a perturbed spacetime with metric $g_{\mu\nu} = \eta_{\mu\nu} + h_{\mu\nu}$, the TTF can be decomposed into two contributions:
\begin{equation}
    \mathcal{T}_r(\vec{X}_e,t_r,\vec{X}_r)=\frac{L_{(0)}}{c}+\frac{1}{c}\Delta_r(\vec{X}_e,t_r,\vec{X}_r)\, ,
    \label{eq:TTFr}
\end{equation}
where $L_{(0)} = \lVert\vec{X}_r - \vec{X}_e\rVert$ is the Euclidean distance between the emission and reception points in flat spacetime, and $\Delta_r$ encodes the reception time delay, induced by the gravitational perturbations.

The pM expansion of the TTF developped in \cite{teyssandierGeneralPostMinkowskianExpansion2008} shows that, at linear order in the metric perturbation, the TTF writes $\mathcal T_r = L_{(0)}/c + \frac{1}{c} \Delta_r^{(1)}(\vec X_e, t_r, \vec{X}_r) + \mathcal O(h^2)$ with\footnotemark
\begin{align*}
    \Delta_r^{(1)}&(\vec{X}_e,t_r,\vec{X}_r)=\\
    &\frac{L_{(0)}}{2}\int_0^1\left[
    h_{00}+2\hat{n}_{(0)}^ih_{0i}+\hat{n}_{(0)}^i\hat{n}_{(0)}^jh_{ij}
    \right]\Big\lvert_{\mathbf{z}(\lambda)}\: d\lambda \, ,
\end{align*}
where $\mathbf{z}(\lambda)$ is the flat spacetime light ray trajectory, i.e. $\mathbf{z}(\lambda)=\left(ct_r-\lambda L_{(0)} , \vec X_r - \lambda L_{(0)}\hat n_{(0)}\right)$.  In the TT gauge, the components $h_{00}$, $h_{0i}$ and $h_{i0}$ vanish and the delay function simplifies to
\begin{equation}
    \Delta_r^{(1)}(\vec{X}_e,t_r,\vec{X}_r)=\frac{L_{(0)}}{2}\hat{n}_{(0)}^i\hat{n}_{(0)}^j\int_0^1  h_{ij} \lvert_{\mathbf{z}(\lambda)}\: d\lambda\, .
\end{equation}
Since the source is assumed to be distant, we employ the plane wave approximation, which means that the GW depends only on $\xi=ct-\hat k\cdot \vec x$, where $\hat{k}$ is the direction of GW propagation. The GW phase along the laser path is then described by
\begin{equation}
    \xi(\lambda)=ct_r-\lambda L_{(0)}-\hat{k}\cdot[\vec{X}_r(t_r)-\lambda L_{(0)}\hat{n}_{(0)}]\, ,
\end{equation}
With this, the first-order time delay becomes
\begin{equation}\label{eq:Deltar}
    \frac{1}{c}\Delta_r^{(1)}=\frac{1}{2c}\frac{\hat{n}_{(0)}^i\hat{n}_{(0)}^j}{1-\hat{k}\cdot\hat{n}_{(0)}}\int_{\xi_e}^{\xi_r}h_{ij}(\xi)\:d\xi.
\end{equation}
This is the well-known expression for the gravitational time delay encountered in systems with stationary spacecraft~\cite{cornishLISAResponseFunction2003, cornishAlternativeDerivationResponse2009}.

\footnotetext{Note that \cite{teyssandierGeneralPostMinkowskianExpansion2008} uses the (+,-,-,-) metric signature. }

\subsubsection{Moving spacecraft}
Let us now consider the spacecraft velocity, to first-order in $\mathcal O(v/c)$. The emission time $t_e = t_r - \mathcal T_r$ is now the solution of an implicit equation
\begin{equation}
    t_r-t_e= \frac{\lVert\vec{X}_r - \vec{X}_e(t_e)\rVert}{c}  +\frac{1}{c}\Delta_r\left(\vec{X}_e(t_e),t_r,\vec{X}_r\right)\, ,
\end{equation}
which we will solve iteratively by using a pM expansion. We introduce a pM expansion of $t_e\approx t_e^{(0)}+t_e^{(1)}+\dots$ and noting that $\Delta_r$ is already a 1pM order term, the previous equation becomes
\begin{widetext}
\begin{align}
    \mathcal T_r = \mathcal T_r^{(0)} +  \mathcal T_r^{(1)} =t_r-t_e^{(0)}-t_e^{(1)} &\approx \frac{\lVert\vec{X}_r - \vec{X}_e(t_e^{(0)}+t_e^{(1)})\rVert}{c} + \frac{1}{c}\Delta_r\left(\vec{X}_e(t_e^{(0)}),t_r,\vec{X}_r\right)  \, , \nonumber\\
    &\approx  \frac{\lVert\vec{X}_r - \vec{X}_e(t_e^{(0)})\rVert}{c} - t_e^{(1)}\vec \beta_e \cdot \frac{\vec X_r - \vec{X}_e(t_e^{(0)})}{\lVert\vec{X}_r - \vec{X}_e(t_e^{(0)})\rVert}+\frac{1}{c}\Delta_r\left(\vec{X}_e(t_e^{(0)}),t_r,\vec{X}_r\right) \, ,\label{eq:TTF_pert}
\end{align}
\end{widetext}
where the approximation means that terms at second pM order are neglected. We have also neglected the spacecraft acceleration in this expansion (i.e. terms of order $\mathcal O(\vec A_e t_e^{(1)})$ and terms that are of the order $\mathcal O(\beta_e^2 t_e^{(1)})$.

At zeroth pM order, the previous equation becomes
\begin{equation}\label{eq:trG0}
   \mathcal T^{(0)}_r =  t_r - t_e^{(0)} = \frac{\lVert\vec{X}_r - \vec{X}_e(t_e^{(0)})\rVert}{c} =\frac{L_{(0)}}{c} \, ,
\end{equation}
which can be solved iteratively or, at first-order in $\beta_e=v_e/c$ to give $\mathcal T^{(0)}_r = D_{(0)}/c + \Delta D /c + \mathcal O(1/c^3)$,  where $D_{(0)} = \lVert\vec{X}_r - \vec{X}_e(t_r)\rVert$ is the instantaneous distance between the emitter and receiver and $\Delta D$ is provided by Eq.~(\ref{eq:Lapprox}).

At first pM order, Eq.~(\ref{eq:TTF_pert}) simply reads
\begin{equation}
\begin{split}
    \mathcal T^{(1)}_r =  - t_e^{(1)} &=  - t_e^{(1)}\vec \beta_e \cdot \frac{\vec X_r - \vec{X}_e(t_e^{(0)})}{\lVert\vec{X}_r - \vec{X}_e(t_e^{(0)})\rVert} \\
    &+\frac{1}{c}\Delta_r\left(\vec{X}_e(t_e^{(0)}),t_r,\vec{X}_r\right) \, ,
\end{split}
\end{equation}
whose solution is  given by
\begin{equation}\label{eq:trG1}
    \mathcal T^{(1)}_r =  \left(1+\vec \beta_e \cdot \hat n^{(0)}\right)\frac{1}{c}\Delta_r\left(\vec{X}_e(t_e^{(0)}),t_r,\vec{X}_r\right)  + \mathcal O(\beta_e^2)\,, 
\end{equation}
with $\hat n^{(0)}$ defined in \eqref{eq:napprox} (and in particular, $\hat n^{(0)}=\hat n^{(0)}+\mathcal O(\beta)$). Note that Eqs. \eqref{eq:trG0} and \eqref{eq:trG1}  coincide with the time delays obtained in Eq.~\eqref{eq:Lapprox} and \eqref{eq:deltat}. 

The fractional frequency shift is defined by $y=\nu_r/\nu_e-1$. It is interesting to decomposed its expression in three contributions
\begin{align}    
    \frac{\nu_r}{\nu_e} = &\left(\left.\frac{d\tau}{dt}\right|_{e}\right) \frac{dt_e}{dt_r} \left(\left.\frac{d\tau}{dt}\right|_{r}\right)^{-1} = \frac{dt_e}{dt_r}+\mathcal O(\beta^2)\, , 
\end{align}
the conversion between proper-time and coordinate time bringing correction at the $\mathcal O(\beta^2)$ in the TT-gauge. As a consequence, the GW contribution to the relative frequency shift is given by
\begin{equation}
    y_\mathrm{GW} = -\frac{d\mathcal{T}^{(1)}_r}{dt_r}\, .
\end{equation}
Hence, we differentiate Eq.~\eqref{eq:trG1} with respect to the reception time $t_r$ to determine the response function. We work to first-order in the spacecraft velocities $\vec{\beta}$, and at 1pM order; terms of order $\mathcal{O}(h^2)$, $\mathcal{O}(\beta^2)$, etc., are neglected. This leads to a tedious calculation since additional non-vanishing derivatives appear once terms linear in the velocities $\vec{\beta}$ are retained. For example, due to the flexing in the arm length over time, we must account for the following derivatives:
\begin{align}
    \frac{1}{c}\frac{dL_{(0)}}{dt_r}&\approx(\vec{\beta}_r-\vec{\beta}_e)\cdot\hat{n}_{(0)}
    \label{eq:dtL}\\
    \frac{1}{c}\frac{d\hat{n}_{(0)}^i}{dt_r}&\approx\frac{1}{L_{(0)}}\left[\beta_r^i-\beta_e^i-\hat{n}_{(0)}\cdot(\vec{\beta}_r-\vec{\beta}_e)\:\hat{n}_{(0)}^i\right]
    \label{eq:dtn0}
\end{align}
Furthermore, the GW phase along a given null geodesic also shifts over time, since the geodesic's boundary points change in time. As we keep $t_r$ and $\vec{X}_r(t_r)$ fixed, we obtain two distinct derivatives for the phase at the boundary points:
\begin{align}
    \frac{1}{c}\frac{d\xi_e}{dt_r}&\approx 1 - (\vec{\beta}_r-\vec{\beta}_e)\cdot\hat{n}_{(0)} - \hat{k}\cdot \vec{\beta}_e
    \label{eq:dtXie}\\
    \frac{1}{c}\frac{d\xi_r}{dt_r}&\approx 1 - \hat{k}\cdot \vec{\beta}_r
    \label{eq:dtXir}
\end{align}
The derivative of the time delay Eq.~\eqref{eq:Deltar}, therefore, contains a term proportional to the time derivative of the geometric factor $\hat{n}_{(0)}^l \hat{n}_{(0)}^m/(1-\hat{k}\cdot\hat{n}_{(0)})$ as well as one proportional to the derivative of the integral $H_{lm}$. Using Eqs.~\eqref{eq:dtL}--\eqref{eq:dtn0}, the former becomes:

\begin{widetext}
\begin{align*}
    \frac{1}{2c}\frac{d}{dt_r}\left(\frac{\hat{n}_{(0)}^l \hat{n}_{(0)}^m}{1-\hat{k}\cdot\hat{n}_{(0)}}\right) H_{lm}=&
    \left(\beta_r^m-\beta_e^m-\hat{n}_{(0)}\cdot(\vec{\beta}_r-\vec{\beta}_e)\:\hat{n}_{(0)}^m\right)
    \frac{\hat{n}_{(0)}^l H_{lm}}{L_{(0)}(1-\hat{k}\cdot\hat{n}_{(0)})}\\
    &+\left[\hat{k}\cdot(\vec{\beta}_r-\vec{\beta}_e)-\hat{n}_{(0)}\cdot(\vec{\beta}_r-\vec{\beta}_e)\,\hat{k}\cdot\hat{n}_{(0)}\right]
    \frac{\hat{n}_{(0)}^l\hat{n}_{(0)}^m H_{lm}}{2L_{(0)}(1-\hat{k}\cdot\hat{n}_{(0)})^2}+\mathcal{O}(\beta^2)\,.
\end{align*}
\end{widetext}

By factoring out the common contribution $(\vec{\beta}_r-\vec{\beta}_e)$ and regrouping the terms proportional to $H_{lm}$, we identify the contraction $\delta_{ij}(\beta^i_r-\beta_e^i)\hat{n}^j_{(1)}$ with the same $\hat{n}_{(1)}$ as we find in Eq.~\eqref{eq:n1}.

Next, we consider the second contribution that arises from the derivative of the metric's integral $H_{lm}$. As a derivative of an integral, this is the value of the metric at the boundary points $\xi_r$ and $\xi_e$ multiplied by the factors given in Eq.~\eqref{eq:dtXir} and \eqref{eq:dtXie}, respectively, to account for the change in variables:

\begin{align*}
    \frac{1}{c}\frac{dH_{lm}}{dt_r}=&h_{lm}(\xi_r)\left[1-\hat{k}\cdot\vec{\beta}_r\right]\\
    &-h_{lm}(\xi_e)\left[1-(\vec{\beta}_r-\vec{\beta}_e)\cdot\hat{n}_{(0)}-\hat{k}\cdot\vec{\beta}_e\right]
\end{align*}

This specifies all the terms present in the derivative of $\frac{1}{c}\Delta_r$. Note that to calculate the full response to the GW, we also need to take the factor $(1+\vec \beta_e \cdot \hat n^{(0)})$ in Eq.~\eqref{eq:trG1} into account. Since we neglect the spacecraft acceleration and higher order terms, we can simply multiply the $t_r$-derivative of $\frac{1}{c}\Delta_r$ by this prefactor. Collecting all contributions mentioned above, yields the fractional frequency shift:

\begin{widetext}
\begin{equation}
    \begin{split}
        y_\mathrm{GW}=-\frac{d\mathcal{T}^{(1)}_r}{dt_r}=&-\frac{1}{2}\frac{\hat{n}_{(0)}^l\hat{n}_{(0)}^m}{1-\hat{k}\cdot\hat{n}_{(0)}}
        \left[
        (1-\vec{\beta}_r\cdot\hat{k}+\hat{n}_{(0)}\cdot\vec{\beta}_e)\:h_{lm}(\xi_r)
        -(1-\vec{\beta}_e\cdot\hat{k}+\hat{n}_{(0)}\cdot(\vec{\beta}_r-2\vec{\beta}_e))\:h_{lm}(\xi_e)
        \right]
        \\
        &-\delta_{ij}(\beta_r^i-\beta_e^i)\hat{n}_{(1)}^j+\mathcal{O}(\beta^2 h)\,,
    \end{split}
    \label{eq:responseTTF}
\end{equation}
\end{widetext}
where we have several contributions dependent on the GW metric $h$ at the boundary points in the first line and the contributions proportional to the metric's integral $H_{lm}$, summarized in $\hat{n}_{(1)}$, in the second line. The expression in Eq.~\eqref{eq:responseTTF} matches the response previously obtained in Sec.~\ref{sec:derivation}, provided in Eq.~\eqref{eq:response}.

\subsection{\label{app:phase} Spacecraft trajectory contributions}
In the previous section, we focuses on the impact of the GW on the propagation of light between the emitter and the receiver (assuming their trajectories to be known). Here, we will include the impact of the GW on the spacecraft motion themselves, as discussed in Section~\ref{sec:trajectory}.

The light-travel time between the two test-masses is given by Eqs.~(\ref{eq:trG0}) and (\ref{eq:trG1}) where $\Delta_r$ is given by Eq.~(\ref{eq:Deltar}). If we now introduce, in the zeroth order term, the decomposition of the spacecraft position obtained in Eq.~(\ref{eq:position}), i.e.
\begin{equation}
    X^i_{r}=X^i_{r,(0)} + X^i_{r(1)}=X^i_{r,(0)} -\tilde H_{r,ij} \beta^j_{r(0)}\, ,
\end{equation}
we get the total first order contribution to the light-travel time given by
\begin{widetext}
\begin{equation}
\begin{split}
\Delta \mathcal T &= \frac{1}{c} \hat n^i_{(0)} \cdot \left(X^i_{r(1)}(t_r)-X^i_{e(1)}(t_e^{(0)})\right)+ \left(1+\vec \beta_e \cdot \hat n^{(0)}\right)\frac{1}{c}\Delta_r\left(\vec{X}_e(t_e^{(0)}),t_r,\vec{X}_r\right)\\
&=\frac{1}{c} \Delta_r - \frac{1}{c}  \hat n^i_{(0)} \cdot\left( \tilde H_{r,ij} \beta^j_{r(0)} -\tilde H_{e,ij} \beta^j_{e(0)}\right) + \beta_e \cdot \hat n^{(0)} \frac{1}{2c}\frac{\hat n^l_{(0)}\hat n^m_{(0)}}{1-\hat k\cdot\hat n_{(0)}}H_{lm}.
\end{split}
\end{equation}
\end{widetext}
Taking the derivative of this expression with respect to $t_r$ yields the expression for the full response, Eq.~\eqref{eq:full_response}.

\section{\label{app:LF} Low-frequency approximation}

In the low-frequency (LF) regime, where the GW wavelength $\lambda$ satisfies $\lambda \gg L_{(0)}$, the detector constellation can be treated as effectively point-like. In this limit, differences of the GW metric evaluated at distinct spacecraft can be expanded to leading order as derivatives:
\begin{equation}
    h_{lm}(\xi_r)-h_{lm}(\xi_e) \stackrel{\text{LF}}{\approx} L_{(0)}(1-\hat{k}\cdot\hat{n}_{(0)})\dot{h}_{lm}(\xi_r),
\end{equation}
where a dot denotes a derivative with respect to the GW phase $\xi$ (with $\xi_e$ and $\xi_r$ the GW phase at emission and reception, respectively). The modified response, Eq.~\eqref{eq:full_response}, also depends on integrals of the GW metric along the null geodesic connecting the spacecraft. In the same limit, these integrals reduce to
\begin{equation}
    H_{lm}=\int_{\xi_e}^{\xi_r}\mathrm{d}\xi \: h_{lm}(\xi) \stackrel{\text{LF}}{\approx} L_{(0)}(1-\hat{k}\cdot\hat{n}_{(0)})h_{lm}(\xi_r).
\end{equation}
Using these expressions, the individual terms in Eq.~\eqref{eq:full_response} can be approximated as
\begin{widetext}
\begin{align}
    \frac{1}{2}\frac{\hat{n}_{(0)}^l\hat{n}_{(0)}^m}{1-\hat{k}\cdot\hat{n}_{(0)}}[h_{lm}(\xi_r)-h_{lm}(\xi_e)]&\stackrel{\text{LF}}{\approx} L_{(0)}\hat{n}_{(0)}^l\hat{n}_{(0)}^m\dot{h}_{lm}(\xi_r),\\
    \frac{1}{2}\frac{\hat{n}_{(0)}^l\hat{n}_{(0)}^m}{1-\hat{k}\cdot\hat{n}_{(0)}}[\vec{\beta}_r^{(0)}\cdot\hat{k}\:h_{lm}(\xi_r)-\vec{\beta}_e^{(0)}\cdot\hat{k}\:h_{lm}(\xi_e)]&\stackrel{\text{LF}}{\approx}\frac{1}{2}\vec{\beta}_r^{(0)}\cdot\hat{k}L_{(0)}\hat{n}_{(0)}^l\hat{n}_{(0)}^m\dot{h}_{lm}(\xi_r)+\Delta\vec{\beta}_{(0)}\cdot\hat{k}\frac{1}{2}\frac{\hat{n}_{(0)}^l\hat{n}_{(0)}^m}{1-\hat{k}\cdot\hat{n}_{(0)}}h_{lm}(\xi_e),\\
    \frac{1}{2}\frac{\hat{n}_{(0)}^l\hat{n}_{(0)}^m}{1-\hat{k}\cdot\hat{n}_{(0)}}[h_{lm}(\xi_r)-h_{lm}(\xi_e)]&\stackrel{\text{LF}}{\approx} 
    -\frac{1}{2}\hat{n}_{(0)}\cdot\vec{\beta}_e^{(0)} L_{(0)}\hat{n}_{(0)}^l\hat{n}_{(0)}^m\dot{h}_{lm}(\xi_r)\\
    -\delta_{ij}(\beta_{r,(0)}^i-\beta_{e,(0)}^i)\hat{n}_{(1)}^j &\stackrel{\text{LF}}{\approx} - \Delta\beta_{(0)}^m n_{(0)}^l h_{lm}(\xi_r) 
    -\frac{1}{2}\Delta \vec{\beta}_{(0)}\cdot \hat{k} \frac{\hat{n}_{(0)}^l\hat{n}_{(0)}^m}{1-\hat{k}\cdot\hat{n}_{(0)}} h_{lm}(\xi_r)\\
    &\qquad+\frac{1}{2}\Delta\vec{\beta}_{(0)}\cdot\hat{n}_{(0)} \hat{n}_{(0)}^l\hat{n}_{(0)}^m h_{lm}(\xi_r)
    +\frac{1}{2}\Delta \vec{\beta}\cdot \hat{n}_{(0)} \frac{\hat{n}_{(0)}^l\hat{n}_{(0)}^m}{1-\hat{k}\cdot\hat{n}_{(0)}} h_{lm}(\xi_r), \nonumber\\
    \left[\beta_{r,(0)}^ih_{ij}(\xi_r)-\beta_{e,(0)}^ih_{ij}(\xi_e)\right] &\stackrel{\text{LF}}{\approx} L_{(0)}(1-\hat{k}\cdot\hat{n}_{(0)})\beta_{r,(0)}^l\hat{n}_{(0)}^m\dot{h}_{lm}(\xi_r)+\Delta\beta^l\hat{n}_{(0)}^mh_{lm}(\xi_r),
\end{align}
where we denote the relative velocity as $\Delta\vec{\beta}\equiv\vec{\beta}_r-\vec{\beta}_e$. Combining the contributions from different terms in the response function, Eq.~\eqref{eq:full_response}, some differences of the metric can be further reduced to derivatives of the metric. This way, the full response can be approximated in the LF limit as:
\begin{equation}
\begin{split}
    y_\mathrm{GW} \stackrel{\text{LF}}{\approx}& 
    -\frac{L_{(0)}}{2}\hat{n}_{(0)}^l\hat{n}_{(0)}^m\dot{h}_{lm}(\xi_r) 
    + \frac{L_{(0)}}{2}\hat{k}\cdot\vec{\beta}_e^{(0)}\hat{n}_{(0)}^l\hat{n}_{(0)}^m\dot{h}_{lm}(\xi_r)
    - \frac{L_{(0)}}{2}\hat{n}_{(0)}\cdot\vec{\beta}_e^{(0)}\hat{n}_{(0)}^l\hat{n}_{(0)}^m\dot{h}_{lm}(\xi_r)\\
    &+ \frac{L_{(0)}}{2}\hat{n}_{(0)}\cdot(\vec{\beta}_r^{(0)}-\vec{\beta}_e^{(0)})\hat{n}_{(0)}^l\hat{n}_{(0)}^m\dot{h}_{lm}(\xi_r)
    +L_{(0)}(1-\hat{k}\cdot\hat{n}_{(0)})\beta_{e,(0)}^l\hat{n}_{(0)}^m\dot{h}_{lm}(\xi_r)\\
    &+\frac{1}{2}\hat{n}_{(0)}\cdot(\vec{\beta}_r^{(0)}-\vec{\beta}_e^{(0)})\hat{n}_{(0)}^l\hat{n}_{(0)}^mh_{lm}(\xi_r).
    \label{eq:response_LF}
\end{split}
\end{equation}

\end{widetext}

\begin{figure*}
    \centering
    \includegraphics[width=\linewidth]{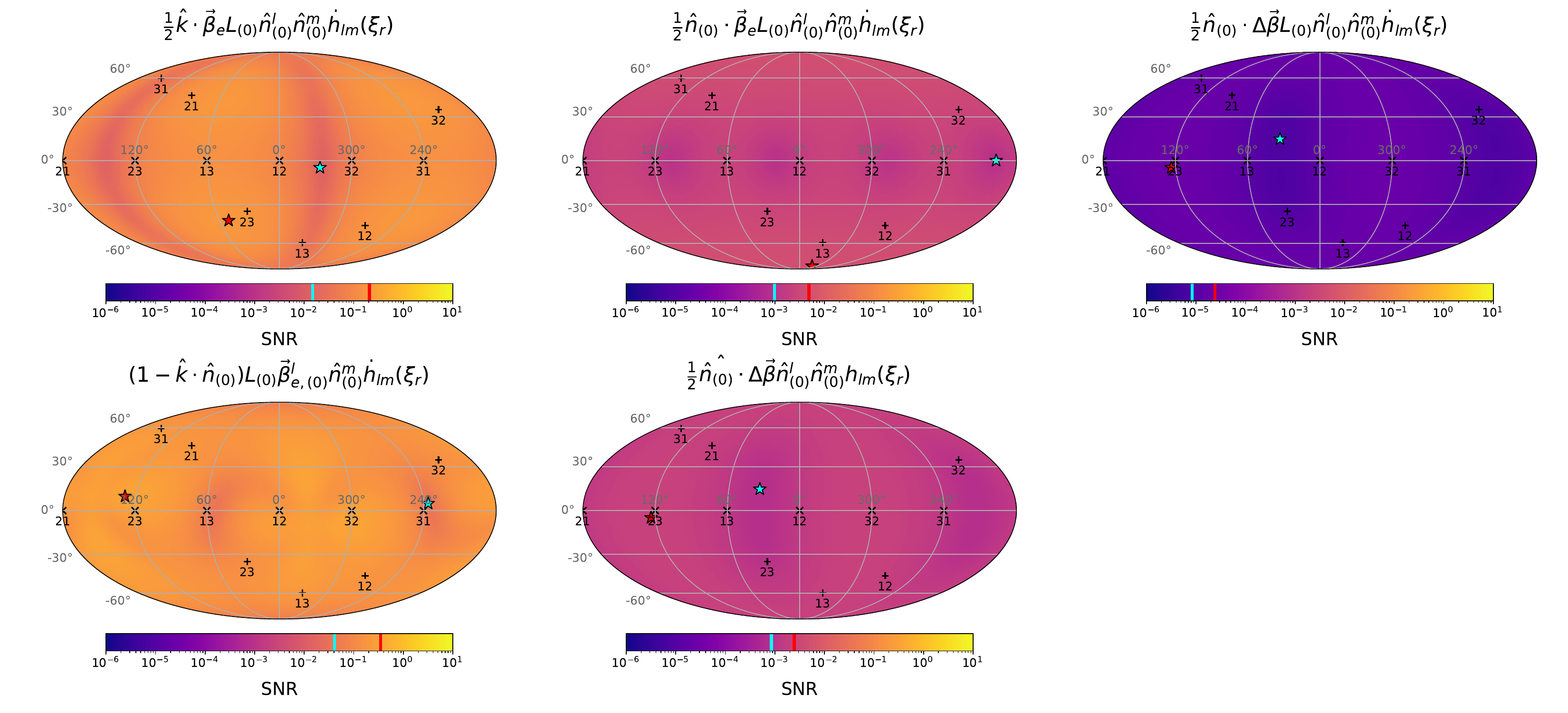}
    \caption{Skymaps of the each correction term in the LF approximation for a source with $M=\SI{e8}{\mathrm{M}_\odot}$. Other parameters are chosen the same as in Fig.~\ref{fig:Skymap_LISAplane}. The reference plane is the LISA plane at time of merger. The labeled crosses ($\times$) indicate the directions of the link vectors $\hat{n}^{(0)}_{IJ}$ and the labeled plus signs ($+$) indicate the direction of the relative velocities $\Delta \vec{\beta}_{IJ}=\vec{\beta}_I-\vec{\beta}_J$. All panels use the same color scale. The red and cyan star indicate the maximum and minimum residual, respectively, for each term. This maximal and minimal SNR value is also indicated in the colorbar for each individual sky-map. In the titles, we use the shorthand $C^{lm} \equiv \hat{n}_{(0)}^l \hat{n}_{(0)}^m / (2[1 - \hat{k} \cdot \hat{n}_{(0)}])$ and $\Delta \beta \equiv\beta_r-\beta_e$. The sky-maps were generated with $N_\mathrm{pix} = 768$ and upsampled using the \texttt{smoothing} function.}
    \label{fig:corrections_LF}
\end{figure*}

To illustrate the sky-position dependence of the individual LF terms, Fig.~\ref{fig:corrections_LF} shows the skymaps of the residuals associated with each term in Eq.~\eqref{eq:response_LF}. The source is a high-mass binary with a total redshifted mass of $M=\SI{e8}{\mathrm{M}_\odot}$. The LF approximation is appropriate for such systems because they accumulate most of their SNR at low frequencies.

\newpage
\section{\label{app:Extra} Complementary material}

\begin{figure*}[t]
    \centering
    \includegraphics[width=\textwidth]{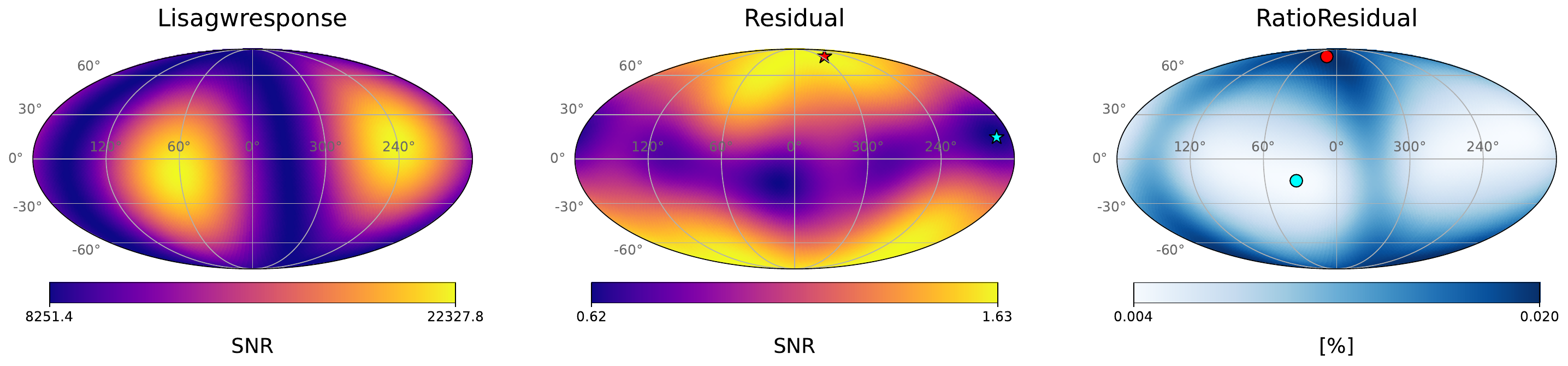}
    \caption{Rotated version of Fig.~\ref{fig:Skymap_LISAplane}. The sky map is rotated such that the reference plane is the equatorial plane. The red and cyan star indicate the maximum and minimum residual, respectively, and the red and cyan dot indicate the maximum and minimum ratio, respectively.}
    \label{fig:Skymap_Equatorial}
\end{figure*}

\begin{figure*}
    \centering
    \includegraphics[width=\textwidth]{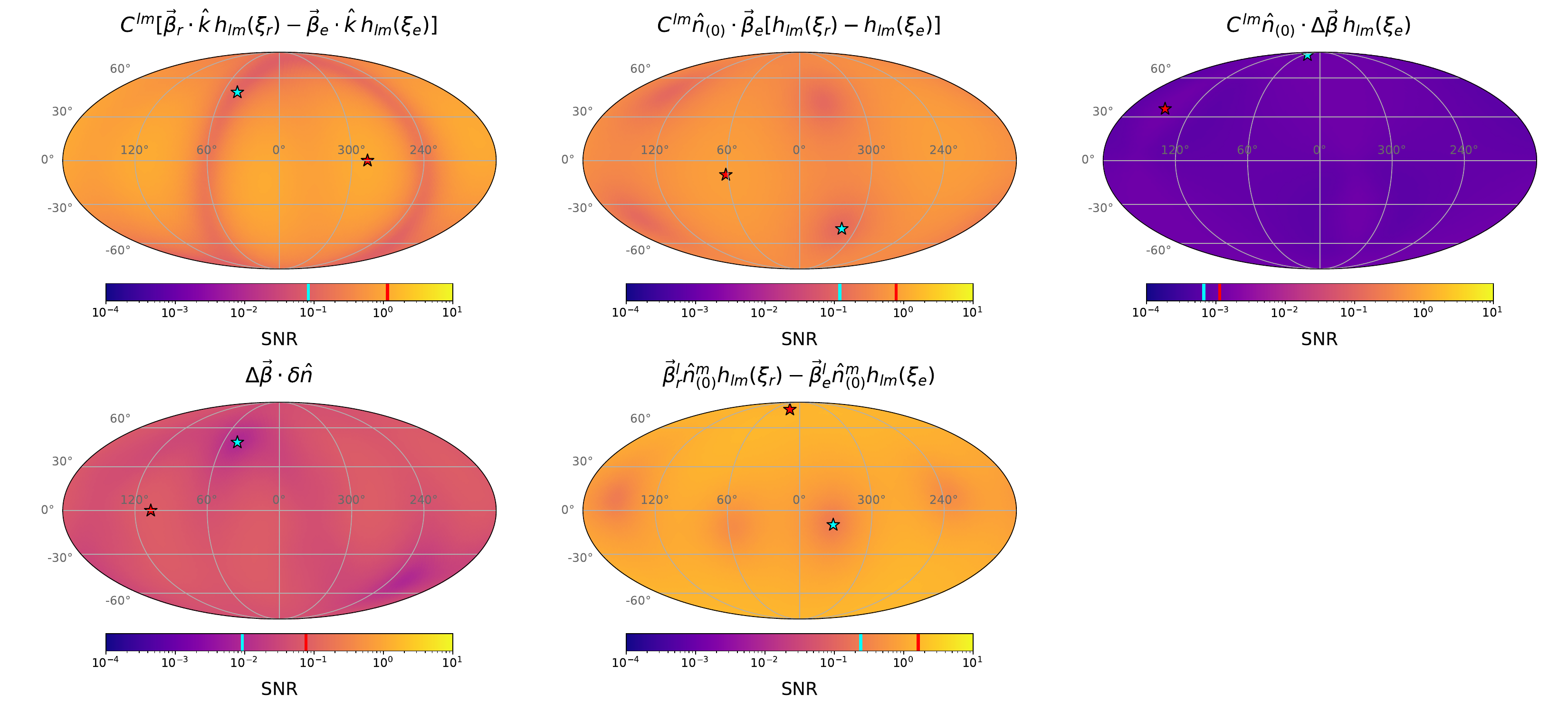}
    \caption{Rotated version of Fig.~\ref{fig:Skymap_Terms_LISAplane}. The sky map is rotated such that the reference plane is the equatorial plane. In titles, we use the abbreviation $C^{lm} \equiv \hat{n}_{(0)}^l \hat{n}_{(0)}^m /(2[1 - \hat{k} \cdot \hat{n}_{(0)}])$ and $\Delta \beta =\beta_r-\beta_e$. The red and cyan star indicate the maximum and minimum residual, respectively, for each term. This maximal and minimal SNR value is also indicated in the colorbar for each individual skymap.}
    \label{fig:Skymap_Terms_Equatorial}
\end{figure*}

\begin{figure*}
    \centering
    \includegraphics[width=\textwidth]{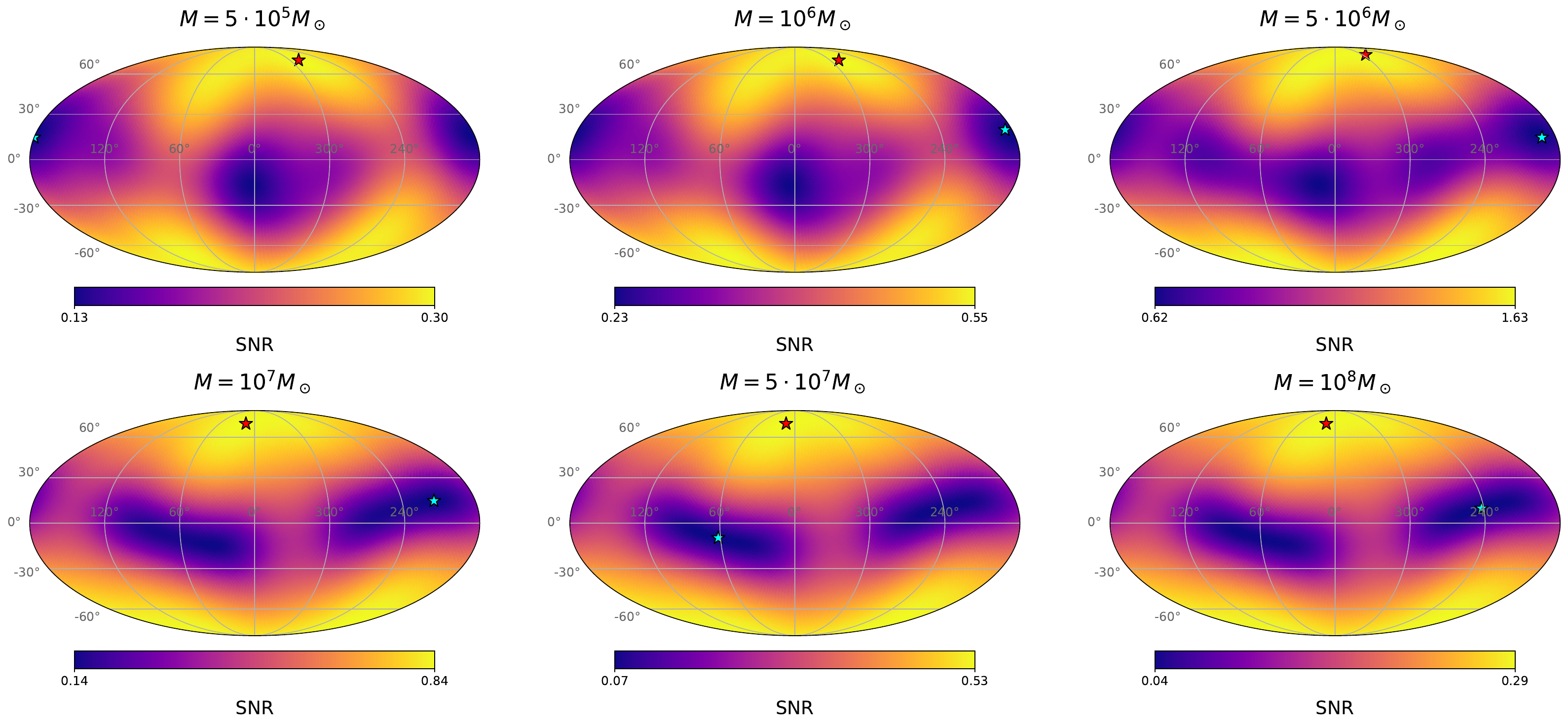}
    \caption{Sky-location dependence of the residual SNR for different total masses. The mass is indicated in the title of each sky-map. Other parameters are chosen the same as Fig.~\ref{fig:Skymap_LISAplane}. The plane of reference is the equatorial plane. The red and cyan star indicate the maximum and minimum residual, respectively.}
    \label{fig:Skymaps_Ms_Res}
\end{figure*}

\begin{figure*}
    \centering
    \includegraphics[width=\textwidth]{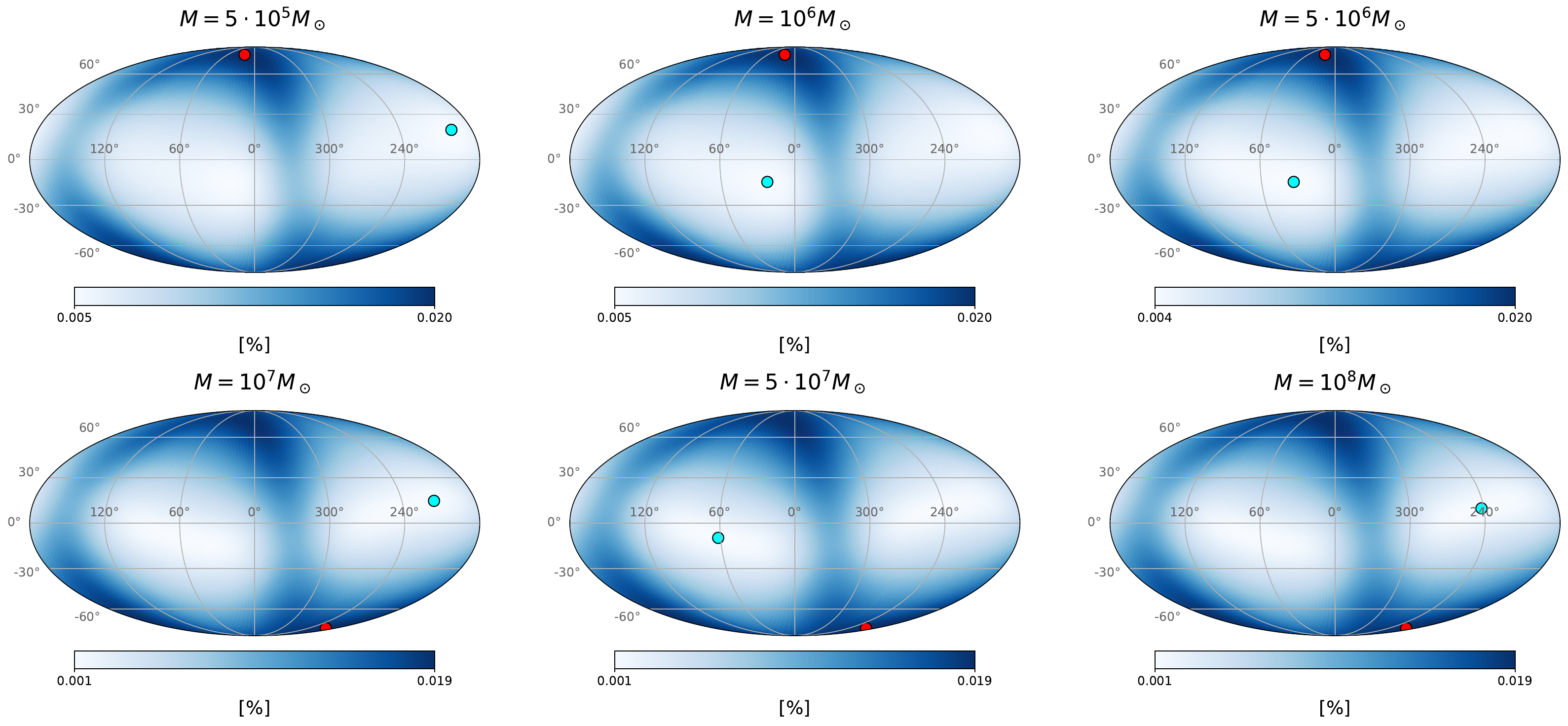}
    \caption{Sky-location dependence of the ratio for different total masses. The mass is indicated in the title of each sky-map. Other parameters are chosen the same as Fig.~\ref{fig:Skymap_LISAplane}. The plane of reference is the equatorial plane. The red and cyan dot indicate the maximum and minimum residual, respectively.}
    \label{fig:Skymaps_Ms_Ratio}
\end{figure*}

In this section, we present Figs.~\ref{fig:Skymap_Equatorial} and \ref{fig:Skymap_Terms_Equatorial}, which are rotated versions of the sky maps shown in Figs.~\ref{fig:Skymap_LISAplane} and \ref{fig:Skymap_Terms_LISAplane}. The maps are displayed in the ICRS equatorial frame rather than in the LISA-plane frame at merger \cite{ariasExtragalacticReferenceSystem1995, lisaddpcconventionsworkinggroupLISARosettaStone2025}. We also show how the sky maps evolve with total mass through a series of maps spanning a range of masses: residuals are given in Fig.~\ref{fig:Skymaps_Ms_Res} and ratios in Fig.~\ref{fig:Skymaps_Ms_Ratio}.

\end{document}